\documentclass{article}

\usepackage{microtype}
\usepackage{multicol}
\usepackage{graphicx}
\usepackage{subcaption}
\usepackage{mdframed}
\usepackage{tabularx}
\usepackage{booktabs}
\usepackage{makecell}

\usepackage{booktabs} 

\usepackage{xurl}
\usepackage{hyperref}

\usepackage[accepted]{icml2026}

\usepackage{amsmath}
\usepackage{amssymb}
\usepackage{mathtools}
\usepackage{amsthm}
\hypersetup{hidelinks}

\usepackage[capitalize,noabbrev]{cleveref}

\theoremstyle{plain}

\theoremstyle{definition}

\theoremstyle{remark}

\usepackage[textsize=tiny]{todonotes}

\icmltitlerunning{Reading Between The Lines: Modeling and Evaluating Behavioral Realism in Legal Simulation}

\begin{document}

\twocolumn[
\icmltitle{Reading Between The Lines: Modeling and Evaluating Behavioral Realism in Legal Simulation}

\icmlsetsymbol{equal}{*}

\begin{icmlauthorlist}
    \icmlauthor{Divya Vetticaden}{equal,stanford}
    \icmlauthor{Arya Gupta}{equal,stanford}
    \icmlcorrespondingauthor{Divya Vetticaden}{divyavetticaden@cs.stanford.edu}
    \icmlcorrespondingauthor{Arya Gupta}{aryagupt@stanford.edu}
    \icmlauthor{Julian Nyarko}{stanford}
    \icmlauthor{Megan Ma}{stanford}
\end{icmlauthorlist}

\icmlaffiliation{stanford}{Stanford University, Stanford, California, USA}

\icmlkeywords{Machine Learning, ICML}

\vskip 0.3in
]
\printAffiliationsAndNotice{\textsuperscript{*} Denotes co-first authorship.}



\begin{abstract}
Deposition training requires attorneys to manage dynamic witness behavior, yet legal-AI evaluations largely focus on factual accuracy, reasoning, or response-level plausibility. We introduce \textsc{WitnessSim}, a deposition simulator driven by controllable legal personas. We use an evaluation framework separating behavioral realism from pedagogical usefulness. We assess realism through adversarial testing, blinded attorney comparison, and analysis of longitudinal behavioral trajectories.  \textsc{WitnessSim} generally maintained plausible behavioral boundaries, and attorneys did not systematically prefer either original testimony or  \textsc{WitnessSim} generated testimony. Pedagogical tests showed that witness behavior changed meaningfully in response to question form and attorney intervention without uniformly collapsing the assigned persona. Together, these results showcase a model of behavioral fidelity in legal simulations, and provide a framework for evaluating its performance.
\end{abstract}

\section{Introduction}
 Depositions are a strategically critical phase of litigation that are important to properly learn. However, junior attorneys typically practice through time-intensive and inconsistent mock exercises, or high stakes participation in real depositions. These limitations are especially acute outside resource-rich legal organizations. Moreover, effective deposition practice requires learning to control nonresponsive testimony, force specificity, identify inconsistencies, manage resistance, and adapt as a witness becomes evasive, defensive, hostile, anxious, or fatigued. These interpersonal and behavioral skills are central to deposition practice but remain largely unaddressed by current legal-AI benchmarks, which primarily evaluate factual accuracy, legal reasoning, or performance on isolated questions. LLM-based simulation offers a promising way to provide attorneys with repeated and configurable opportunities to practice these skills. Thus, we introduce  \textsc{WitnessSim}, a dynamic persona based deposition-training system designed to simulate witnesses with distinct witness behavioral archetypes.  \textsc{WitnessSim} instantiates a witness from underlying case materials and, as questioning progresses, updates an interpretable state representing multiple psychological and behavioral dimensions. These updates are driven by features of the attorney’s questions, including pressure, topic sensitivity, and question form, and are used to condition the witness’s subsequent responses. This structure is intended to make behavioral evolution interpretable and inspectable rather than leaving it to latent and imprecise conversational states of underlying language models.

A central challenge, however, is evaluation: what makes a simulated witness effective for legal training? Producing plausible individual responses is not enough. A witness may appear superficially realistic while failing to respond meaningfully to pressure, contradiction, or changes in questioning strategy. Conversely, a simulator may generate a strategically difficult interaction without producing behavior that resembles natural testimony. Evaluation must, therefore, consider both whether the witness behaves credibly and whether the interaction gives attorneys opportunities to practice legally meaningful skills. Thus, we evaluate \textsc{WitnessSim} along two complementary dimensions: behavioral realism and pedagogical usefulness.

The behavioral-realism layer asks whether generated testimony could plausibly have been produced by a human witness and whether the witness maintains a coherent behavioral identity throughout the interaction. It combines adversarial testing, expert assessment of contextual plausibility, and affective evaluation of how the witness’s demeanor changes over time. The pedagogical-usefulness layer asks whether  \textsc{WitnessSim} produces the kinds of difficult interactions attorneys are expected to recognize and manage in practice. We ground this evaluation in a review of legal training materials and feedback from experienced attorneys, translating recurring training objectives into tests of whether witnesses respond appropriately to pressure, question form, sensitive topics, contradictions, and archetype-specific questioning strategies. Finally, we evaluate this framework using a randomly sampled corpus of 300 deposition and trial transcripts drawn from 1,169 candidate transcripts in the National Prescription Opiate Litigation (Case No. 1:17-MD-2804), a federal multidistrict litigation consolidating cases against opioid manufacturers, distributors, and pharmacies, alongside synthetic interactions generated with  \textsc{WitnessSim}.

Our contributions are as follows: First, we introduce  \textsc{WitnessSim}, a controllable, state-based system for simulating dynamic witness behavior during deposition questioning. Second, we develop an evaluation framework that distinguishes behavioral realism from pedagogical usefulness. The framework operationalizes realism through adversarial testing, expert assessment of contextual plausibility, and a novel, exploratory emotion-vector analysis of trajectory-level affective dynamics, while evaluating pedagogical usefulness through legally grounded tests of whether witness behavior creates recognizable challenges and responds meaningfully to attorney intervention. Together, these contributions frame deposition simulation as a dynamic, multi-turn problem in which simulation quality depends not only on plausible individual responses, but also on how behavior develops across an interaction and questioning.

\section{Related Work}

This work lies at the intersection of three areas: LLM-based social and persona simulation, affective and conversational behavior modeling, and simulation for professional training.

\textbf{LLM-Based Social and Persona Simulation.}
Recent work has shown that LLM agents can generate coherent individual and collective social behavior. Generative Agents demonstrated that agents could form routines, respond to events, and sustain open-ended social interactions, while \citep{he2026collective} extended this work to larger populations that reproduced community-level patterns such as homophily \citep{park2023generativeagents}. In these settings, fidelity is defined broadly as coherent and recognizably human behavior rather than agreement with a particular person's responses. More tightly grounded approaches instead attempt to reproduce individual attitudes and behavioral patterns from interviews or surveys. \citep{park2024thousand} construct agents representing specific people, while \citep{wang2025personality} evaluate personality simulations through psychometric reliability, structural validity, heterogeneity, and behavioral prediction. Together, these studies treat personality as an underlying construct that should produce consistent behavior across contexts.

Persona-based simulation occupies a middle ground between open-ended social modeling and replication of a specific individual. Benchmarks such as PersonaGym and Eval4Sim evaluate whether behavior remains plausible and consistent with an assigned persona across changing contexts, using dimensions such as persona adherence, identity consistency, and conversational naturalness \citep{bao2026eval4sim, samuel2025personagym}. \citep{lutz2025prompt} further show that how underlying traits are specified can materially affect generated behavior, diversity, alignment, and stereotyping. This work establishes that persona construction and trait definition shape simulation output, but largely evaluates response-level consistency rather than behavior under sustained adversarial interaction.

Within law, AgentCourt \citep{chen2025agentcourt} and SimCourt \citep{zhang2025simcourt} model courtroom proceedings, emphasizing procedural structure, legal reasoning, and agent performance, while benchmarks such as LegalBench \citep{guha2023legalbench} evaluate capabilities including rule recall, issue spotting, and legal application.  \textsc{WitnessSim} builds on this work by adding a complementary unit and dimension of evaluation: the behavior of an individual witness across sustained questioning. Rather than assessing whether a model can reproduce a legal process or arrive at a legally appropriate answer,  \textsc{WitnessSim} asks whether a simulated witness maintains a coherent persona, responds plausibly to adversarial pressure, and creates behaviorally meaningful challenges for attorneys. It therefore extends existing legal-AI simulation and benchmarking from replicating and measuring procedural and doctrinal competence toward the interpersonal dynamics that shape legal practice.

\textbf{Affective and Conversational Behavior Modeling.}
Affective-computing research studies how emotion, personality, and interpersonal behavior can be represented and measured computationally \citep{picard1997affective}. Conversation-level emotion-recognition models increasingly treat affect as dependent on interaction history: DialogueGCN \citep{ghosal2019dialoguegcn} models within- and between-speaker dependencies, EmoRoBERTa \citep{kim2021emoberta} incorporates speaker identity, and DialogXL  \citep{shen2021dialogxl} maintains longer-range utterance context . These approaches motivate evaluating witness behavior across dialogue rather than treating each answer independently.

Complementary work examines how psychological characteristics shape generated language. PsychAdapter \citep{vu2026psychadapter} conditions language models on continuous personality, demographic, and mental-health profiles, demonstrating that psychological representations can be used to control generated behavior. \citep{sofroniew2026emotion} identify directional representations of emotion concepts in model activations and show that they track emotional contexts and can influence generation. We adapt this emotion-vector method as an exploratory measure of trajectory-level affective dynamics. More broadly,  \textsc{WitnessSim} treats interpersonal behavior as something that must be both generated and evaluated as it unfolds through interaction.

\textbf{LLM Simulation for Professional Training.}
Behavioral fidelity alone does not make a simulator useful for training; the interaction must also create opportunities to practice the intended skills. LLM role-play has supported training in negotiation, conflict resolution, counseling, and medicine by enabling repeated practice without continuous access to instructors or standardized participants. The AI Partner--AI Mentor framework \citep{yang2024socialskill} combines experiential role-play with individualized feedback, while Rehearsal \citep{louie2024rehearsal} allows users to practice difficult conflicts and explore alternative strategies. \citep{noh2024personalities}  further show that Big Five conditioning produces systematic differences in negotiation behavior, including agreement, exploitation, toxicity, and language use, demonstrating that personality conditioning can create distinct interactional challenges.

Accordingly, professional-training systems are evaluated through more than dialogue quality, including trainee strategy, expert judgments, perceived usefulness, and learning outcomes. Rehearsal, for example, compares conflict-resolution strategy use following simulation-based and lecture-based instruction. Within legal education, \citep{rolf2025contract} uses AI-generated contractual language in an experiential business-law exercise evaluated through engagement, critical thinking, and doctrinal learning. More directly, AI-Assisted Moot Courts \citep{zhang2026mootcourts} separates realism from pedagogical usefulness: its realism evaluation asks whether simulated justice questions resemble authentic questioning, while its pedagogical evaluation asks whether they surface issues and weaknesses advocates should practice addressing.

Witness simulation requires the same distinction but a different training target. Whereas an oral-argument simulator should expose advocates to legal and argumentative weaknesses, a witness simulator must create behavioral challenges that attorneys can recognize and manage.  \textsc{WitnessSim} therefore evaluates both whether witness behavior remains plausible and coherent and whether it responds meaningfully to legally relevant intervention.

Taken together, prior work shows that LLMs can simulate plausible social agents, express assigned personas, model affective behavior, and support professional role-play. However, it leaves open how to evaluate a witness simulator whose quality depends on behavior across a sustained adversarial interaction. Plausible individual responses do not establish longitudinal behavioral fidelity, and apparent realism does not establish training relevance.

 \textsc{WitnessSim} addresses this gap by treating deposition simulation as both a behavioral-realism and pedagogical-evaluation problem. We evaluate whether simulated witnesses maintain plausible behavioral identities over time and whether they create recognizable examination challenges that change meaningfully in response to attorney questioning. This connects persona simulation and conversational behavior modeling with the specific demands of legal training.

\section{Methods}
\subsection{ \textsc{WitnessSim}}
 \textsc{WitnessSim} is a deposition simulator built around a continuous dynamical system that models how a witness's psychological state evolves under adversarial questioning. The content of each answer is conditioned on a six state vector that is updated deterministically from features of the attorney's question, and then ultimately informs an LLM for generation. 

\subsubsection{State Space and Archetypes}
\label{sec:state-space}
At deposition turn \(t\), the witness is represented by the state vector
\begin{equation}
    \mathbf{y}_t
    =
    \left(
        C_t,K_t,A_t,V_t,R_t,P_t
    \right)
    \in [0,1]^6,
\end{equation}
where \(C\), \(K\), \(A\), \(V\), \(R\), and \(P\) denote
\emph{composure}, \emph{knowledge}, \emph{agreeableness},
\emph{verbosity}, \emph{rigidity}, and \emph{performance},
respectively. Composure loosely corresponds to inverse neuroticism,
agreeableness retains its conventional interpretation, and rigidity
corresponds approximately to inverse openness. Knowledge, verbosity, and
performance capture deposition-specific behaviors that are not represented
cleanly by the conventional Big Five dimensions.

We began with fourteen expert-informed candidate archetypes. Each candidate
was represented as a point in the six-dimensional state space, and pairwise
cosine similarity and angular separation were used to identify candidates
whose behavioral profiles were not meaningfully distinct. This process
produced the ten archetypes used in the present study:
\emph{combative}, \emph{cooperative}, \emph{defensive},
\emph{dogmatic}, \emph{inventive}, \emph{loquacious},
\emph{nervous}, \emph{neutral}, \emph{overconfident}, and
\emph{overprepared}.

The complete retained archetype vectors and their approximate Big Five
projections are reported in
Tables~\ref{tab:archetype-six-state} and
\ref{tab:archetype-big-five}, with the full pairwise comparison reported
in Table~\ref{tab:archetype-pairwise} of
Appendix~\ref{app:archetype-comparison}.

Each retained archetype is associated with an attractor state
\(\mathbf{y}_0\in[0,1]^6\). This attractor represents the witness's baseline
behavior: in the absence of sustained pressure, the dynamic state gradually
returns toward \(\mathbf{y}_0\). The attractor is initialized from case
materials during persona generation and remains fixed throughout the
simulated deposition.

\paragraph{Illustrative distinction: Inventive versus Loquacious.}
While existing personality frameworks such as MBTI and the Big 5 also categorize behavior, they are subpar for modelling contexts. An example of the benefit of the six-state system, that retains a separate knowledge state is particularly clear for the Inventive and Loquacious archetypes. In this case, their
six-state attractors are
\begin{equation}
\begin{aligned}
    \mathbf{y}_{0}^{\mathrm{inv}}
        &= (0.70,\,0.20,\,0.60,\,0.85,\,0.25,\,0.60),\\
    \mathbf{y}_{0}^{\mathrm{loq}}
        &= (0.60,\,0.70,\,0.60,\,0.95,\,0.25,\,0.55).
\end{aligned}
\label{eq:inventive-loquacious-vectors}
\end{equation}
Both archetypes are relatively composed, moderately agreeable, highly
verbal, non-rigid, and similarly prepared. Their central distinction is,
therefore, the knowledge coordinate: the Inventive witness fills gaps through
fabrication or improvisation, whereas the Loquacious witness possesses the
relevant information but communicates it at excessive length.

For two archetype vectors \(\mathbf{u}\) and \(\mathbf{v}\), similarity is
measured by
\begin{equation}
    \operatorname{cos}(\mathbf{u},\mathbf{v})
    =
    \frac{\mathbf{u}^{\top}\mathbf{v}}
         {\|\mathbf{u}\|_2\|\mathbf{v}\|_2},
    \qquad
    \theta(\mathbf{u},\mathbf{v})
    =
    \arccos\!\left(
        \operatorname{cos}(\mathbf{u},\mathbf{v})
    \right).
\label{eq:archetype-cosine}
\end{equation}
In the full six-dimensional state space, the pair has cosine similarity
approximately \(0.944\), corresponding to an angular separation of about
\(19^\circ\).

To illustrate what is lost under a Big Five representation, consider the
approximate projection
\begin{equation}
    \Pi(C,K,A,V,R,P)
    =
    (1-C,\ A,\ V,\ 1-R,\ P),
\label{eq:big-five-projection}
\end{equation}
whose coordinates correspond approximately to neuroticism,
agreeableness, extraversion, openness, and conscientiousness. The knowledge
coordinate \(K\) has no direct analog and is therefore omitted. Under this
projection,
\begin{equation}
\begin{aligned}
    \Pi\!\left(\mathbf{y}_{0}^{\mathrm{inv}}\right)
        &= (0.30,\,0.60,\,0.85,\,0.75,\,0.60),\\
    \Pi\!\left(\mathbf{y}_{0}^{\mathrm{loq}}\right)
        &= (0.40,\,0.60,\,0.95,\,0.75,\,0.55).
\end{aligned}
\end{equation}
Their projected cosine similarity increases to approximately \(0.996\), and
their angular separation falls to approximately \(5^\circ\). Thus, the Big
Five projection treats the two witnesses as almost identical, whereas the
six-state representation preserves the behaviorally important distinction
between not knowing and merely talking too much.

\subsubsection{Question Encoding}
\label{sec:question-encoding}
Before each state update, the attorney's question \(q_t\) is encoded into a
pressure score and a topic-sensitivity score.

The pressure score \(p_t\in[0,1]\) is computed from prespecified linguistic
markers:
\begin{equation}
    p_t
    =
    \operatorname{clip}_{[0,1]}
    \left(
        0.1
        +0.2\,n_{\mathrm{high}}(q_t)
        +0.1\,n_{\mathrm{med}}(q_t)
    \right),
\label{eq:pressure-score}
\end{equation}
where \(n_{\mathrm{high}}\) and \(n_{\mathrm{med}}\) count high- and
medium-pressure markers appearing in the question. The additive structure
allows multiple adversarial cues to accumulate, while clipping prevents the
result from leaving the interpretable interval \([0,1]\). The complete marker dictionary and implementation details are reported in
Table~\ref{tab:pressure-markers} and
Appendix~\ref{app:question-encoding-details}.
Each persona also contains a set of preregistered sensitive topics
\(\mathcal{T}\). For topic \(\tau\), token overlap with the current question
is measured by the Jaccard coefficient
\begin{equation}
    J_t(\tau)
    =
    \frac{
        \left|
            \operatorname{tok}(q_t)
            \cap
            \operatorname{tok}(\tau)
        \right|
    }{
        \left|
            \operatorname{tok}(q_t)
            \cup
            \operatorname{tok}(\tau)
        \right|
    }.
\label{eq:jaccard-sensitivity}
\end{equation}
The resulting sensitivity score is
\begin{equation}
    s_t
    =
    \max_{\tau\in\mathcal{T}}
    \left\{
        \sigma_\tau J_t(\tau)
        \mathbb{I}\!\left[J_t(\tau)>0.15\right]
    \right\},
\label{eq:sensitivity-score}
\end{equation}
where \(\sigma_\tau\in[0,1]\) is the topic's intrinsic sensitivity weight.
The threshold suppresses incidental lexical overlap, while multiplication by
\(\sigma_\tau\) distinguishes merely relevant topics from topics that are
especially destabilizing for the witness. The numerical coefficients and behavioral rationale for the six state
updates are reported in
Table~\ref{tab:state-dynamics-justification}, with illustrative fatigue
values in Table~\ref{tab:fatigue-schedule} of
Appendix~\ref{app:state-dynamics}. Additional implementation details for sensitive-topic construction and
tokenization are provided in
Appendix~\ref{app:sensitive-topic-tokenization}.

Pressure, sensitivity, and question structure are combined into a composite
stress signal
\begin{equation}
    \xi_t
    =
    0.5p_t
    +0.3s_t
    +0.2\mathbb{I}_{\mathrm{lead},t},
\label{eq:composite-stress}
\end{equation}
where \(\mathbb{I}_{\mathrm{lead},t}=1\) when \(q_t\) is classified as a
leading question. Raw pressure is the dominant contribution, while topic
sensitivity and leading form add distinct sources of stress.

\subsubsection{State Dynamics}
\label{sec:state-dynamics}

The six state dimensions evolve through coupled discrete-time updates that
combine an immediate response to the current question with gradual
mean-reversion toward the witness's archetype-specific baseline. Pressure,
topic sensitivity, and question form affect dimensions differently:
composure decreases under stress; sensitive questioning can degrade recall;
pressure reduces cooperation and increases rigidity; sensitive topics can
increase verbosity while leading questions constrain it; and performance
deteriorates increasingly as fatigue accumulates over the deposition.

The resulting update can be written generally as
\begin{equation}
    \mathbf{y}_{t+1}
    =
    \operatorname{clip}_{[0,1]}
    \left(
        \mathbf{y}_t +
        F(\mathbf{y}_t,p_t,s_t,\xi_t,t;\mathbf{y}_0)
    \right),
\end{equation}
where \(F\) denotes the dimension-specific state dynamics and
\(\mathbf{y}_0\) is the archetype attractor. This structure allows witness
behavior to respond dynamically to questioning while preventing short-term
interactions from immediately overriding the underlying persona. The full
dimension-specific update equations, numerical coefficients, and behavioral
rationales are provided in Appendix~\ref{app:state-dynamics}.

\subsubsection{Derived Behavioral Scores}
\label{sec:derived-scores}

Four interpretable behavioral summaries are computed from the current state:
\begin{equation}
\begin{aligned}
    \operatorname{consistency}_t
        &=
        0.5K_t + 0.4R_t
        \\
        &\quad
        - 0.1(1-K_t)(1-R_t),
        \\[3pt]
    \operatorname{evasion}_t
        &=
        0.5(1-K_t)
        + 0.3\lvert 2V_t-1\rvert
        \\
        &\quad
        + 0.2P_t,
        \\[3pt]
    \operatorname{realism}_t
        &=
        \max\Bigl(
            0,\,
            1 - 0.3\lvert 2C_t-1\rvert
        \\
        &\qquad
            - 0.3\lvert 2K_t-1\rvert
            - 2\max(0,P_t-0.8)
        \Bigr),
        \\[3pt]
    \operatorname{adversarial}_t
        &=
        0.4(1-K_t)
        + 0.3(1-A_t)
        \\
        &\quad
        + 0.2R_t
        + 0.1(1-C_t).
\end{aligned}
\label{eq:derived-behavioral-scores}
\end{equation}

Consistency is driven primarily by knowing the relevant facts and maintaining a stable account. The interaction term imposes an additional penalty when knowledge and rigidity are simultaneously low, corresponding to a witness who neither knows the account nor adheres consistently to it.

Evasion increases when the witness lacks knowledge, communicates at either extreme of verbosity, or is sufficiently prepared to respond strategically. Both extreme terseness and extreme verbosity can obscure a direct answer.

Realism penalizes extreme composure and extreme recall, as well as performance levels sufficiently high to resemble coaching rather than ordinary human
behavior. Adversarial behavior combines unreliability, non-cooperation, entrenchment, and composure loss, with the largest weights placed on knowledge gaps and unwillingness to cooperate. Additional motivation for the construction and relative weighting of these
four behavioral summaries is provided in
Table~\ref{tab:derived-score-rationale} of
Appendix~\ref{app:derived-score-rationale}.

\subsubsection{Event Detection}

A small set of discrete behavioral events is fired when threshold
conditions on the state vector are met (Table~\ref{tab:events}). The
triggered events, along with the current state, are appended to the
LLM system prompt at generation time, anchoring the surface dialogue
in the underlying dynamics rather than allowing the LLM to drift.

\begin{table}[t]
\centering
\small
\begin{tabular}{p{1.8cm}p{5cm}}
\toprule
\textbf{Event} & \textbf{Condition} \\
\midrule
Witness rattled     & $p_t > 0.58 \wedge (C_t < 0.38 \vee \Delta C_t < -0.05)$ \\
Witness combative   & $A_t < 0.28 \wedge R_t > 0.60$ \\
Attorney interrupts & $p_t > 0.45 \wedge |q_t| \leq 8\ \text{words} \wedge \mathrm{type}(q_t) \in \{\text{closed, leading}\}$ \\
Witness talks over  & $R_t > 0.72 \wedge V_t > 0.58 \wedge p_t > 0.35$ \\
Personality shift   & $\|\mathbf{s}_t - \mathbf{s}_0\|_\infty > 0.35$ \\
\bottomrule
\end{tabular}
\caption{Threshold conditions on the state vector for discrete
behavioral events. Triggered events are passed to the LLM system
prompt to ground generated dialogue in the current mathematical state.}
\label{tab:events}

\end{table}

\subsubsection{Generation Loop}

For each deposition turn,  \textsc{WitnessSim} (i) encodes the attorney question
into $(p_t, s_t, \xi_t)$, (ii) updates the state vector and per-topic recall quality
$\rho$ (Appendix~\ref{app:recall-quality}), (iii) computes the four derived scores and evaluates the event
predicates, and (iv) prompts the underlying LLM with the persona,
current state, fired events, and conversation history to generate the
witness's answer. The state vector and event flags are  recorded at every
turn, yielding the trajectory data analyzed in
our experimental section.

\section{Evaluation}\label{sec:evaluation_framework}

\subsubsection{Framework and Setup}
We evaluate  \textsc{WitnessSim} along two complementary dimensions: behavioral realism and pedagogical usefulness. Following the two-layer evaluation structure introduced by \citep{zhang2026mootcourts}, behavioral realism asks whether generated testimony resembles plausible human witness behavior, while pedagogical usefulness asks whether the simulator creates recognizable examination challenges that respond meaningfully to attorney intervention. These objectives are related but distinct: a witness may produce fluent testimony while remaining overly compliant or behaviorally static, whereas a strategically difficult simulation may still fail to resemble plausible human testimony.

We randomly sampled 300 deposition and trial transcripts from 1,169 candidate transcripts in the UCSF Industry Documents Library’s opioid-litigation corpus. For each sampled witness,  \textsc{WitnessSim} was grounded in the corresponding transcript and, where available, supplementary case materials. Before evaluation, each simulated witness completed a warm-up phase using up to ten preceding questions from the original transcript. Questions that would reveal the later test topic were excluded, ensuring that all evaluation prompts began from a grounded but unexposed conversational state. Additional sampling and warm-up details appear in
Appendix~\ref{app:corpus-warmup}.
Judge-derived outcomes were compared against independent human annotations conducted by the authors. Full rubrics and reliability analyses are provided in Appendix ~\ref{app:judge-reliability}.

\subsection{Behavioral Realism}
We assess behavioral realism through three complementary evaluations: adversarial testing, blinded expert comparison with authentic testimony, and longitudinal analysis of affective and behavioral dynamics. Together, these evaluations examine whether  \textsc{WitnessSim} maintains plausible behavioral boundaries under manipulation, produces testimony that practicing attorneys consider contextually believable, and exhibits coherent behavior across longer interactions.

\subsubsection{Adversarial Behavioral Robustness}
\label{sec:adversarial-robustness}

We designed four adversarial attacks targeting common failure modes in persona-based LLM agents. AC1 tests whether a
Cooperative witness becomes overly compliant when directly asked to accept
personal blame; AC2 extends this pressure across five increasingly accusatory
leading premises to test whether cooperation remains bounded rather than
collapsing into either immediate resistance or unrestricted agreement. AC3
tests whether an Evasive witness resists selectively by comparing responses
to three trivial and three case-sensitive questions from the same starting
state. Finally, AC4 asks the identical question five consecutive times across
the Cooperative, Evasive, and Combative archetypes to test consistency,
repetition awareness, and preservation of archetype-specific reactions.
Responses were scored using prespecified behavioral rubrics, with human
annotation used to validate the automated measures. Table~\ref{tab:adversarial-overview}
summarizes the primary outcomes; full prompts, scoring rules, reliability
analyses, and statistical tests are reported in
Appendix~\ref{app:adversarial-full}.

\begin{table}[t]
\centering
\caption{Adversarial tests and primary outcomes.}
\label{tab:adversarial-overview}
\small
\renewcommand{\arraystretch}{1.15}
\begin{tabularx}{\columnwidth}{@{}p{0.10\columnwidth}XX@{}}
\toprule
\textbf{Test} & \textbf{Adversarial pressure} & \textbf{Primary result} \\
\midrule

AC1 &
Direct pressure on a Cooperative witness to accept personal blame &
$99.0\%$ preserved a self-protective boundary while remaining in character \\

AC2 &
Escalating chain of increasingly accusatory leading premises &
All witnesses resisted before accepting the complete chain; $72.7\%$
resisted within the intended interval \\

AC3 &
Trivial versus sensitive questions posed to an Evasive witness &
$86.0\%$ showed greater evasion on sensitive than trivial material \\

AC4 &
Five consecutive repetitions of the same question across three archetypes &
Responses remained consistent, avoided mechanical duplication, and showed
the expected archetype ordering in irritation onset \\

\bottomrule
\end{tabularx}
\end{table}

Across the four attacks,  \textsc{WitnessSim} generally preserved the intended
behavioral boundary rather than collapsing into a generic failure mode.
In AC1, 297/300 Cooperative witnesses avoided full admission while remaining
in character. In AC2, no witness accepted the complete five-premise chain,
although $27.3\%$ resisted earlier than intended. Evasive witnesses showed
substantially greater resistance to sensitive than trivial material
($r_{\mathrm{rb}}=.984$, $p<.001$), indicating selective rather than blanket
evasion. Under repetition, irritation emerged in the prespecified order
(Combative, Evasive, Cooperative; Page's $L=3730$, $p<.0001$), while a
deterministic similarity check found no exact or near-verbatim duplication
across the 900 evaluated sequences. These results suggest that the simulator
remains responsive to adversarial questioning while largely preserving the
behavioral distinctions encoded by its assigned archetypes.

\subsubsection{Plausibility}
\label{sec:plausibility}

We next evaluated whether \textsc{WitnessSim} reaches a minimum threshold of
contextual plausibility through blinded pairwise judgments. Evaluators
received brief case background, the immediately preceding deposition
dialogue, the attorney's next four questions, and two alternative
four-response sequences: one authentic and one generated by \textsc{WitnessSim},
randomly assigned to Transcript A or B. They selected which sequence was
more plausible as real witness testimony in context, with options to select
either sequence, a tie, or neither.

Three evaluators completed the task. A practicing associate and a senior arbitration counsel
each reviewed the same initial 50-item set. A senior practicing litigator
reviewed a tuned 49-item version. The senior arbitration counsel had an engineering background, had also
previously collaborated in legal simulation development, and was therefore, substantially more
familiar with the system and with artifacts characteristic of LLM-generated
text than the other evaluators. We accordingly report
each evaluator separately rather than pooling their judgments.
\begin{table}[t]
\centering
\caption{Blinded judgments of contextual plausibility.}
\label{tab:expert-plausibility}
\scriptsize
\setlength{\tabcolsep}{2.5pt}
\renewcommand{\arraystretch}{1.15}

\begin{tabularx}{\columnwidth}{
@{}
>{\raggedright\arraybackslash}p{0.27\columnwidth}
c
>{\centering\arraybackslash}X
>{\centering\arraybackslash}X
>{\centering\arraybackslash}X
>{\centering\arraybackslash}X
@{}
}
\toprule
\textbf{Evaluator} &
\textbf{$n$} &
\makecell{\textbf{Witness}\\\textbf{Sim}} &
\textbf{Original} &
\textbf{Tie} &
\textbf{Neither} \\
\midrule

Associate
& 50
& \makecell{15\\(30.0\%)}
& \makecell{10\\(20.0\%)}
& \makecell{20\\(40.0\%)}
& \makecell{5\\(10.0\%)} \\

Senior litigator
& 49
& \makecell{16\\(32.7\%)}
& \makecell{14\\(28.6\%)}
& \makecell{11\\(22.4\%)}
& \makecell{8\\(16.3\%)} \\

\makecell[l]{Senior Arbitration\\Counsel}
& 50
& \makecell{0\\(0.0\%)}
& \makecell{45\\(90.0\%)}
& \makecell{5\\(10.0\%)}
& \makecell{0\\(0.0\%)} \\

\bottomrule
\end{tabularx}
\end{table}

The associate and senior litigator did not systematically prefer authentic
testimony: the associate selected  \textsc{WitnessSim} in 30.0\% of exchanges versus
20.0\% for the original, with 40.0\% ties, while the senior litigator
selected  \textsc{WitnessSim} in 32.7\% versus 28.6\% for the original, with 22.4\%
ties. In contrast, the senior arbitration counsel identified the authentic sequence
in 45/50 exchanges and marked the remaining five as ties. However, it is important to note that this evaluator has an engineering background, has been working on AI transformation at their firm, and has been actively involved in deposition simulation development efforts.  Given this
evaluator's direct familiarity with and experience evaluating
LLM-generated legal text, the divergence suggests that artifacts may remain
detectable to reviewers specifically attuned to model-generated behavior.
The attorney evaluations, nevertheless, support the narrower conclusion that
 \textsc{WitnessSim} frequently falls within the perceived plausibility range of
authentic multi-turn deposition testimony. Descriptions of instructions, evaluator
background, instrument versioning, and additional results are reported in
Appendix~\ref{app:plausibility}.

\subsubsection{Trajectory-Level Behavioral Analysis}
\label{sec:trajectory-analysis}

As a complementary proof of concept, we examine behavioral structure across
longer, multi-turn interactions using emotion-vector trajectories. Unlike
response-level evaluations, this analysis asks whether simulated testimony
captures the persistence and change in behavior that emerge over sustained
questioning. We use externally computed emotion-vector projections as an
exploratory proxy for this longitudinal structure; full construction and
preprocessing details appear in Appendix~\ref{app:trajectory}.

\begin{figure}[t]
    \centering
    \includegraphics[width=\columnwidth]
    {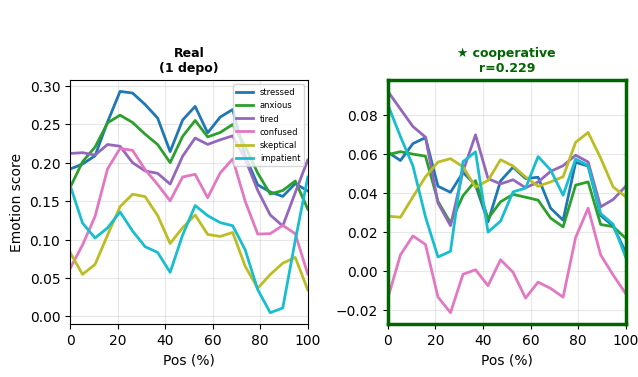}
    \caption{
    Representative emotion-vector trajectories for a real deposition and its
    best-fitting simulated counterpart
    (Jeffrey Kilper, cooperative, $r=.229$).
    }
    \label{fig:arc-comparison}
\end{figure}

Real and simulated depositions exhibited modest but above-chance temporal
alignment. The mean trajectory correlation across emotion dimensions and
simulated archetypes was $r=.148$, exceeding all 1,000 temporally permuted
comparisons ($p<.001$). Figure~\ref{fig:arc-comparison} shows a representative
real--synthetic pair, with several aligned rises and declines but a visibly
smaller dynamic range in the simulated trajectory.

\begin{figure}[t]
    \centering
    \includegraphics[width=\columnwidth]
    {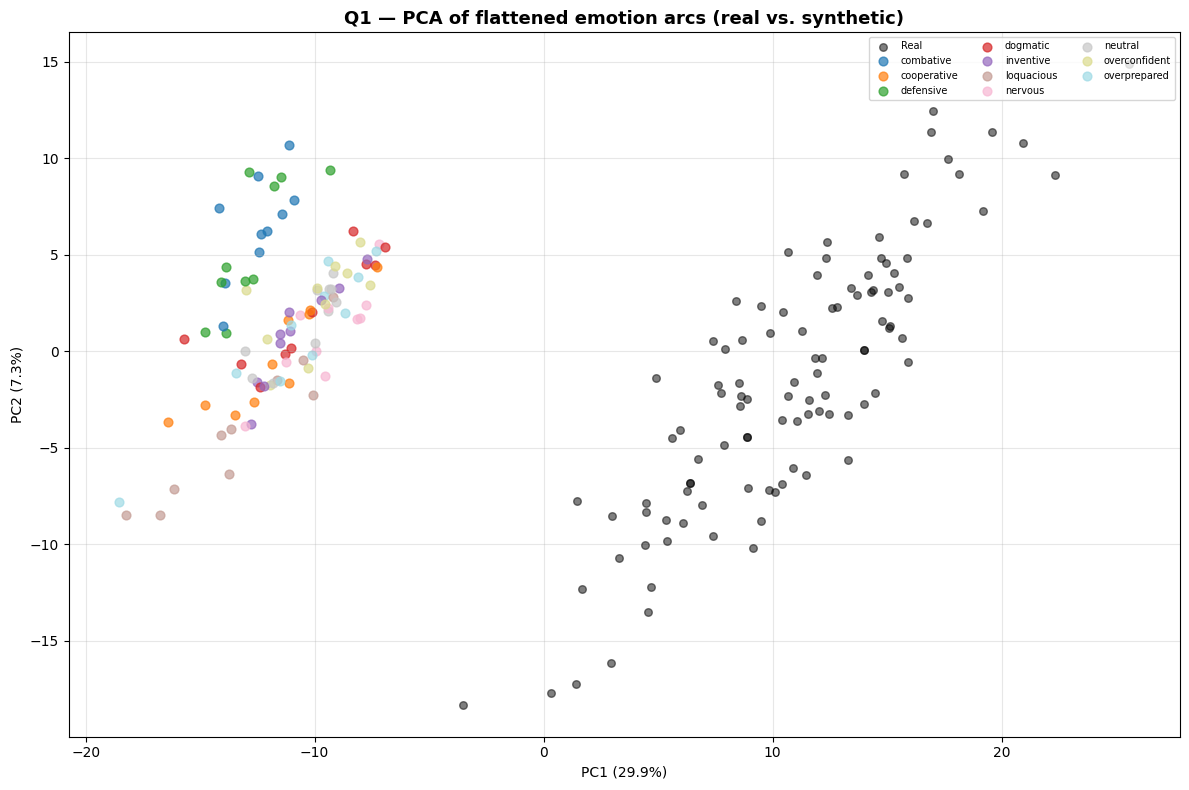}
    \caption{
    PCA of flattened emotion-vector trajectories. Real transcripts are shown
    in gray and synthetic transcripts are colored by archetype.
    }
    \label{fig:trajectory-pca}
\end{figure}

The analysis also revealed systematic fidelity gaps. Synthetic trajectories
were smoother and more compressed than real trajectories, and PCA of the
trajectory representations separated real and simulated transcripts along
the first principal component, which explained 29.9\% of the variance
(Figure~\ref{fig:trajectory-pca}).

These results illustrate why behavioral fidelity cannot be assessed from
individual responses alone. \textsc{WitnessSim} captures some shared temporal
structure, but its smoother, more compressed trajectories indicate remaining
gaps in longer-horizon behavioral realism. Emotion-vector trajectories offer
one exploratory approach to measuring these differences, although the measure
requires further validation.

\subsection{Pedagogical Usefulness}
\label{sec:pedagogical}

The pedagogical-usefulness evaluation asks whether  \textsc{WitnessSim} produces recognizable witness-management challenges and whether those behaviors respond meaningfully to legally relevant questioning techniques. This evaluation does not measure attorney learning or skill transfer; rather, it assesses whether the simulator generates interactions that support the practice of such skills.

We reviewed deposition, cross-examination, trial-advocacy, and witness-control materials
\citep{howard2010witnesscontrol,caldwell2018wreckingball,
tanford1994control,pozner2018cross,maccarthy2007cross,
melvin2015trialadvocacy,miller2021evasive,kautzman2010difficult}
and translated recurring training objectives into candidate evaluation tasks. An experienced litigator then selected four tasks judged most relevant for attorney practice. Table~\ref{tab:pedagogical-summary} summarizes the training target and primary result for each test.

\begin{table}[t]
\centering
\caption{Pedagogical evaluation tasks and headline results.}
\label{tab:pedagogical-summary}
\small
\renewcommand{\arraystretch}{1.15}
\begin{tabularx}{\columnwidth}{@{}p{0.29\columnwidth}X@{}}
\toprule
\textbf{Test} & \textbf{Training target and primary result} \\
\midrule

\textbf{Question Form}
(Cooperative)
&
Adapting question form to control the scope and character of testimony. Open and clarifying questions elicited longer, generally more qualified responses, while closed, leading, and high-pressure questions produced substantially shorter answers. \\

\textbf{Evasive Pin-Down}
(Evasive)
&
Forcing specificity from a witness who avoids commitment. The complete evasion-to-control trajectory occurred in $93.0\%$ of contexts; $33.1\%$ of substantive final answers still retained hedging or qualification. \\

\textbf{Runaway Witness}
(Loquacious)
&
Redirecting lengthy or unfocused testimony. The intended unfocused-to-focused trajectory occurred in $59.3\%$ of contexts, while $40.3\%$ were already focused from the outset; only $0.3\%$ never resolved to a focused answer. \\

\textbf{Hostile Witness}
(Combative)
&
Obtaining usable testimony without eliminating the witness's adversarial demeanor. Substantive control was reached in $97.7\%$ of contexts, and $97.3\%$ of controlled answers retained markers of hostility. \\

\bottomrule
\end{tabularx}
\end{table}

Across the four tests,  \textsc{WitnessSim} generally responded to legally meaningful
attorney intervention while preserving the behavioral challenge associated
with the assigned archetype. Question form produced substantial changes in
testimony: for example, open questions elicited an average of 175.9 words,
compared with 86.5 for closed and 68.4 for high-pressure questions. More
targeted interventions also altered witness responsiveness. In the Evasive
Pin-Down test, the intended evasion-to-control trajectory occurred in
$93.0\%$ of contexts, while $33.1\%$ of substantive final answers still
retained hedging or qualification. Similarly, $97.7\%$ of combative witnesses
reached substantive control, yet $97.3\%$ of those controlled answers retained
markers of hostility. These results suggest that effective questioning can
change the witness's responsiveness without simply collapsing the persona into
generic cooperation.

The loquacious archetype showed a somewhat different pattern. Across the
three mutually exclusive trajectory outcomes, the targeted runaway-to-focused
trajectory emerged in $59.3\%$ of contexts, while $40.3\%$ were already
focused from the outset despite producing characteristically lengthy responses;
only $0.3\%$ exhibited the challenge without ultimately resolving to a focused
answer. Thus, loquacity often manifested as verbosity without loss of focus,
and persistent failure to regain focus once the targeted challenge appeared
was rare. 

Taken together, these results provide evidence that  \textsc{WitnessSim} can generate
behaviorally distinct, legally relevant practice scenarios in which attorney
interventions produce meaningful changes in witness behavior without uniformly
erasing the underlying archetype. Full protocols, scoring criteria, reliability
analyses, and statistical results are reported in Appendix~\ref{app:pedagogical-details} .
\section{Discussion}

\label{sec:discussion}

Across our evaluations, \textsc{WitnessSim} generally produced behavior that
was plausible, responsive to pedagogically meaningful attorney intervention, and
consistent with the behavioral challenge assigned to the witness. At the same
time, the trajectory analysis suggested remaining gaps in longer-horizon
fidelity: simulated testimony exhibited some shared temporal structure with
real testimony but was smoother, more compressed, and distinguishable in the
trajectory representation. These findings suggest that behavioral realism is
not a single property. A simulator may produce convincing individual responses
and useful reactions to intervention while still failing to reproduce the full
variation of human behavior across a sustained interaction.

This distinction becomes increasingly important as LLMs are used to simulate
people and social interaction. Many of the behaviors that matter in
professional settings are inherently multi-turn: rapport
changes, strategies shift, and people differ
in their attitudes, personalities, emotional responses, and behavioral
tendencies. What constitutes a realistic response, therefore, depends not only
on the immediate prompt or prior conversational history, but also on the
characteristics of the person (a) being simulated and how those characteristics
interact with the unfolding exchange. Evaluating isolated responses can miss
these dynamics. High-fidelity human simulation will require methods that
assess not only whether individual responses appear plausible, but also
whether behavior remains coherent with the simulated person and develops,
persists, and changes in plausible ways across an interaction.

This perspective may also matter for how AI is incorporated into professional
training. As people increasingly offload cognitive and professional tasks to
AI systems, concerns about deskilling and the erosion of human capabilities
are becoming more salient. Simulation-based learning offers a different use of
the same technology: rather than replacing the exercise of professional
judgment, AI can be used to create repeated opportunities to practice and
develop it. In litigation, for
example, attorneys may be able to encounter a wider range of difficult witness
behaviors, experiment with different questioning strategies, and receive
practice that would otherwise depend on expensive or difficult-to-scale human
role-play. The value of such systems, however, depends on the quality of the
behavior they simulate. Training against an agent that is unrealistically
compliant, rigid, or behaviorally static may teach the wrong lessons rather
than strengthen the intended skills.

Our evaluation, therefore, establishes only one part of the case for
simulation-based training: whether the simulated interaction contains the
behavioral challenges and responses that make meaningful practice possible.
The next step is to test whether those properties translate into actual
learning. Controlled studies with trainees and practitioners could examine
whether repeated interaction with behaviorally realistic simulations improves
questioning strategy, adaptation to difficult witnesses, or transfer to new
scenarios, and whether differences in simulation fidelity affect those
outcomes. Connecting measures of behavioral realism to downstream learning
would move evaluation beyond asking whether a simulated person appears
plausible towards asking whether the simulation is sufficiently faithful to be
useful.

\section*{Limitations and Future Work}

There are a few limitations present in our work. First, the evaluation
corpus is drawn from a single opioid-litigation context. Broader evaluation across different cases would be a strong next step to determine more robust generalizability. Furthermore, the emotion-vector analysis should be understood as a proof of
concept rather than a validated measure of witness affect. Future studies
should compare these representations with human annotations of demeanor for a more robust analysis. More broadly, although simulation-based learning may offer a way to use AI to strengthen rather than replace professional judgment, our study does not test whether repeated use of \textsc{WitnessSim} mitigates deskilling or translates into measurable improvements in attorney learning. Finally, controlled studies with trainees, instructors, and practicing attorneys should measure learning and performance with and without the use of  \textsc{WitnessSim} to more robustly determine usefulness in the context of adoption.
\section*{Conclusion}
We introduced \textsc{WitnessSim}, a controllable, state-based deposition
simulator, together with an evaluation framework that separates behavioral
realism from pedagogical usefulness. Across adversarial tests, blinded
plausibility judgments, and legally grounded examination tasks,
\textsc{WitnessSim} generally maintained meaningful behavioral distinctions
while responding to changes in attorney questioning. At the same time,
trajectory-level analysis revealed systematic gaps: simulated testimony was
smoother, more compressed, and distinguishable from real testimony over
longer interactions. These results illustrate why response-level plausibility
alone is insufficient for evaluating interactive simulations. More broadly, this work frames behavioral fidelity as a multi-turn evaluation
problem. A useful professional simulator must do more than produce plausible
individual responses: it should maintain coherent behavioral boundaries,
respond meaningfully to intervention, and exhibit realistic change across an
interaction. Although our evaluation does not establish attorney learning or
full equivalence to human witnesses, it provides a framework for measuring
these properties separately and identifying where simulation fidelity
succeeds or breaks down. As LLM-based training and simulation become increasingly common, evaluating how behavior develops and changes across an interaction will be critical to developing high-fidelity, useful simulations.

\section*{Impact Statement}
This paper presents work whose goal is to advance the field of Machine Learning. As with any generative system, models trained on legal data carry inherent risks, including the potential to produce inaccurate, biased, or hallucinated outputs that may not reliably reflect statutes, case law, or jurisdictional nuance. In the legal domain, where errors can affect due process, client outcomes, and access to justice, such limitations warrant particular caution, and we emphasize that the methods presented here are intended to support, not substitute for, the judgment of qualified legal professionals.

\section*{Acknowledgements}

We thank the team at Three Crowns for their contributions to the development
of \textsc{WitnessSim} and for bringing their practical expertise
to the design and evaluation of the system. We are especially grateful to
Hugh Carlson, CEO of Three Crowns, and Nicholas Jampol, Partner at Davis
Wright Tremaine LLP, for serving as expert annotators and contributing their time
and expertise.

\bibliography{example_paper}
\bibliographystyle{icml2026}

\newpage
\appendix
\onecolumn

\section*{Appendix}


\section{Archetype Representation and Big Five Comparison}
\label{app:archetype-comparison}

The simulator began from fourteen expert-informed candidate archetypes.
After examining their locations in the six-dimensional state space,
ten retained archetypes were used in the present evaluation.
Table~\ref{tab:archetype-six-state} reports their complete six-state
attractor vectors.

\begin{table}[h]
\centering
\caption{Six-state attractor vectors for the ten retained witness archetypes.
The dimensions are composure ($C$), knowledge ($K$), agreeableness ($A$),
verbosity ($V$), rigidity ($R$), and performance ($P$).}
\label{tab:archetype-six-state}
\small
\renewcommand{\arraystretch}{1.12}
\begin{tabular}{@{}lcccccc@{}}
\toprule
\textbf{Archetype} & $\mathbf{C}$ & $\mathbf{K}$ & $\mathbf{A}$ &
$\mathbf{V}$ & $\mathbf{R}$ & $\mathbf{P}$ \\
\midrule
Combative      & 0.25 & 0.55 & 0.15 & 0.75 & 0.85 & 0.50 \\
Cooperative    & 0.80 & 0.85 & 0.90 & 0.50 & 0.20 & 0.65 \\
Defensive      & 0.55 & 0.60 & 0.30 & 0.30 & 0.75 & 0.60 \\
Dogmatic       & 0.80 & 0.75 & 0.50 & 0.20 & 0.95 & 0.70 \\
Inventive      & 0.70 & 0.20 & 0.60 & 0.85 & 0.25 & 0.60 \\
Loquacious     & 0.60 & 0.70 & 0.60 & 0.95 & 0.25 & 0.55 \\
Nervous        & 0.20 & 0.45 & 0.55 & 0.75 & 0.40 & 0.25 \\
Neutral        & 0.55 & 0.55 & 0.55 & 0.55 & 0.50 & 0.55 \\
Overconfident  & 0.90 & 0.65 & 0.75 & 0.65 & 0.80 & 0.90 \\
Overprepared   & 0.80 & 0.90 & 0.60 & 0.20 & 0.80 & 0.95 \\
\bottomrule
\end{tabular}
\end{table}

For comparison, we construct the approximate Big Five projection used in the
archetype discussion in Section~\ref{sec:state-space}:
\[
\Pi(C,K,A,V,R,P)
=
(1-C,\ A,\ V,\ 1-R,\ P),
\]
corresponding approximately to neuroticism ($N$), agreeableness ($A$),
extraversion ($E$), openness ($O$), and conscientiousness
($C_{\mathrm{bf}}$). The knowledge coordinate $K$ is omitted because it
has no direct analog in this approximation.

\begin{table}[h]
\centering
\caption{Approximate Big Five projection of the ten retained archetypes.}
\label{tab:archetype-big-five}
\small
\renewcommand{\arraystretch}{1.12}
\begin{tabular}{@{}lccccc@{}}
\toprule
\textbf{Archetype} &
$\mathbf{N}$ &
$\mathbf{A}$ &
$\mathbf{E}$ &
$\mathbf{O}$ &
$\mathbf{C_{\mathrm{bf}}}$ \\
\midrule
Combative      & 0.75 & 0.15 & 0.75 & 0.15 & 0.50 \\
Cooperative    & 0.20 & 0.90 & 0.50 & 0.80 & 0.65 \\
Defensive      & 0.45 & 0.30 & 0.30 & 0.25 & 0.60 \\
Dogmatic       & 0.20 & 0.50 & 0.20 & 0.05 & 0.70 \\
Inventive      & 0.30 & 0.60 & 0.85 & 0.75 & 0.60 \\
Loquacious     & 0.40 & 0.60 & 0.95 & 0.75 & 0.55 \\
Nervous        & 0.80 & 0.55 & 0.75 & 0.60 & 0.25 \\
Neutral        & 0.45 & 0.55 & 0.55 & 0.50 & 0.55 \\
Overconfident  & 0.10 & 0.75 & 0.65 & 0.20 & 0.90 \\
Overprepared   & 0.20 & 0.60 & 0.20 & 0.20 & 0.95 \\
\bottomrule
\end{tabular}
\end{table}

To examine how this projection changes the geometry among archetypes,
we computed cosine similarity and angular separation for all
$\binom{10}{2}=45$ retained-archetype pairs in both spaces.
Table~\ref{tab:archetype-pairwise} reports the complete comparison.
We define
\[
\Delta\theta
=
\theta_{R^5}-\theta_{R^6}.
\]
Thus, negative values indicate pairs that become more similar under the
Big Five projection, whereas positive values indicate pairs that become
more separated.

Importantly, the projection does not uniformly compress all pairwise
distances. Rather, its limitation for the present application is that
dropping $K$ can collapse distinctions for which knowledge is the
behaviorally important differentiating coordinate. The
Inventive--Loquacious pair provides the clearest example: its angular
separation falls from approximately $19.2^\circ$ in the six-state space
to $5.2^\circ$ after the Big Five projection.

\begin{table}[p]
\centering
\caption{Pairwise archetype similarity in the six-state space ($R^6$)
and approximate Big Five space ($R^5$).
$\Delta\theta=\theta_{R^5}-\theta_{R^6}$; negative values indicate
compression under the Big Five projection.}
\label{tab:archetype-pairwise}
\scriptsize
\setlength{\tabcolsep}{5pt}
\renewcommand{\arraystretch}{1.05}
\begin{tabular}{@{}lccccc@{}}
\toprule
\textbf{Pair} &
$\mathbf{\cos_{R^6}}$ &
$\boldsymbol{\theta_{R^6}}$ &
$\mathbf{\cos_{R^5}}$ &
$\boldsymbol{\theta_{R^5}}$ &
$\boldsymbol{\Delta\theta}$ \\
\midrule
Neutral / Overconfident & 0.990 & 8.0$^\circ$ & 0.890 & 27.1$^\circ$ & +19.1$^\circ$ \\
Defensive / Dogmatic & 0.989 & 8.4$^\circ$ & 0.900 & 25.8$^\circ$ & +17.4$^\circ$ \\
Nervous / Overconfident & 0.868 & 29.7$^\circ$ & 0.703 & 45.4$^\circ$ & +15.6$^\circ$ \\
Combative / Dogmatic & 0.847 & 32.1$^\circ$ & 0.678 & 47.3$^\circ$ & +15.2$^\circ$ \\
Combative / Overprepared & 0.818 & 35.1$^\circ$ & 0.639 & 50.3$^\circ$ & +15.2$^\circ$ \\
Combative / Overconfident & 0.871 & 29.4$^\circ$ & 0.714 & 44.5$^\circ$ & +15.0$^\circ$ \\
Defensive / Overconfident & 0.961 & 16.1$^\circ$ & 0.866 & 30.0$^\circ$ & +14.0$^\circ$ \\
Defensive / Overprepared & 0.977 & 12.4$^\circ$ & 0.901 & 25.8$^\circ$ & +13.4$^\circ$ \\
Neutral / Overprepared & 0.940 & 20.0$^\circ$ & 0.841 & 32.8$^\circ$ & +12.8$^\circ$ \\
Dogmatic / Neutral & 0.930 & 21.5$^\circ$ & 0.837 & 33.2$^\circ$ & +11.7$^\circ$ \\
Nervous / Overprepared & 0.754 & 41.0$^\circ$ & 0.611 & 52.3$^\circ$ & +11.3$^\circ$ \\
Dogmatic / Nervous & 0.754 & 41.0$^\circ$ & 0.627 & 51.1$^\circ$ & +10.1$^\circ$ \\
Loquacious / Overprepared & 0.822 & 34.7$^\circ$ & 0.732 & 42.9$^\circ$ & +8.2$^\circ$ \\
Loquacious / Overconfident & 0.915 & 23.8$^\circ$ & 0.853 & 31.5$^\circ$ & +7.6$^\circ$ \\
Cooperative / Overprepared & 0.902 & 25.6$^\circ$ & 0.843 & 32.5$^\circ$ & +7.0$^\circ$ \\
Combative / Cooperative & 0.710 & 44.8$^\circ$ & 0.631 & 50.9$^\circ$ & +6.2$^\circ$ \\
Dogmatic / Loquacious & 0.792 & 37.6$^\circ$ & 0.724 & 43.6$^\circ$ & +6.0$^\circ$ \\
Cooperative / Overconfident & 0.928 & 21.9$^\circ$ & 0.884 & 27.9$^\circ$ & +6.0$^\circ$ \\
Inventive / Overconfident & 0.916 & 23.6$^\circ$ & 0.876 & 28.8$^\circ$ & +5.2$^\circ$ \\
Cooperative / Dogmatic & 0.857 & 31.0$^\circ$ & 0.814 & 35.6$^\circ$ & +4.5$^\circ$ \\
Combative / Neutral & 0.888 & 27.3$^\circ$ & 0.851 & 31.7$^\circ$ & +4.3$^\circ$ \\
Overconfident / Overprepared & 0.958 & 16.6$^\circ$ & 0.936 & 20.6$^\circ$ & +4.0$^\circ$ \\
Combative / Defensive & 0.909 & 24.6$^\circ$ & 0.884 & 27.9$^\circ$ & +3.3$^\circ$ \\
Cooperative / Nervous & 0.845 & 32.3$^\circ$ & 0.819 & 35.0$^\circ$ & +2.7$^\circ$ \\
Loquacious / Nervous & 0.945 & 19.1$^\circ$ & 0.929 & 21.8$^\circ$ & +2.6$^\circ$ \\
Cooperative / Loquacious & 0.934 & 20.9$^\circ$ & 0.923 & 22.7$^\circ$ & +1.8$^\circ$ \\
Dogmatic / Overconfident & 0.954 & 17.5$^\circ$ & 0.946 & 19.0$^\circ$ & +1.5$^\circ$ \\
Cooperative / Defensive & 0.846 & 32.2$^\circ$ & 0.834 & 33.5$^\circ$ & +1.3$^\circ$ \\
Combative / Loquacious & 0.836 & 33.3$^\circ$ & 0.827 & 34.2$^\circ$ & +1.0$^\circ$ \\
Cooperative / Neutral & 0.947 & 18.8$^\circ$ & 0.943 & 19.4$^\circ$ & +0.6$^\circ$ \\
Inventive / Overprepared & 0.776 & 39.1$^\circ$ & 0.770 & 39.7$^\circ$ & +0.6$^\circ$ \\
Dogmatic / Inventive & 0.758 & 40.7$^\circ$ & 0.752 & 41.2$^\circ$ & +0.5$^\circ$ \\
Defensive / Neutral & 0.945 & 19.0$^\circ$ & 0.944 & 19.3$^\circ$ & +0.3$^\circ$ \\
Combative / Nervous & 0.880 & 28.3$^\circ$ & 0.882 & 28.1$^\circ$ & -0.3$^\circ$ \\
Dogmatic / Overprepared & 0.985 & 10.1$^\circ$ & 0.989 & 8.4$^\circ$ & -1.7$^\circ$ \\
Combative / Inventive & 0.771 & 39.6$^\circ$ & 0.791 & 37.7$^\circ$ & -1.9$^\circ$ \\
Nervous / Neutral & 0.921 & 22.9$^\circ$ & 0.934 & 20.9$^\circ$ & -2.0$^\circ$ \\
Inventive / Nervous & 0.880 & 28.4$^\circ$ & 0.899 & 25.9$^\circ$ & -2.5$^\circ$ \\
Defensive / Loquacious & 0.829 & 34.1$^\circ$ & 0.857 & 31.1$^\circ$ & -3.0$^\circ$ \\
Loquacious / Neutral & 0.955 & 17.3$^\circ$ & 0.969 & 14.2$^\circ$ & -3.1$^\circ$ \\
Defensive / Nervous & 0.796 & 37.2$^\circ$ & 0.843 & 32.5$^\circ$ & -4.7$^\circ$ \\
Defensive / Inventive & 0.785 & 38.3$^\circ$ & 0.861 & 30.6$^\circ$ & -7.7$^\circ$ \\
Inventive / Neutral & 0.923 & 22.7$^\circ$ & 0.970 & 14.0$^\circ$ & -8.6$^\circ$ \\
Cooperative / Inventive & 0.881 & 28.2$^\circ$ & 0.947 & 18.8$^\circ$ & -9.4$^\circ$ \\
Inventive / Loquacious & 0.944 & 19.2$^\circ$ & 0.996 & 5.2$^\circ$ & -14.0$^\circ$ \\
\bottomrule
\end{tabular}
\end{table}


\section{Question-Encoding Details}
\label{app:question-encoding-details}

\subsection{Pressure-Marker Dictionary}
\label{app:pressure-markers}

The pressure score uses a baseline of $0.10$ and prespecified
high- and medium-pressure linguistic markers.
Table~\ref{tab:pressure-markers} reports the complete dictionary.

\begin{table}[h]
\centering
\caption{Prespecified linguistic markers used to construct the pressure score.}
\label{tab:pressure-markers}
\small
\renewcommand{\arraystretch}{1.2}
\begin{tabularx}{\textwidth}{@{}p{0.18\textwidth}p{0.15\textwidth}X@{}}
\toprule
\textbf{Component} & \textbf{Contribution} & \textbf{Markers / interpretation} \\
\midrule

Baseline &
$0.10$ &
Applied to every question before marker-specific increments. \\

High-pressure marker &
$+0.20$ per match &
\texttt{isn't it true};
\texttt{you're lying};
\texttt{how do you explain};
\texttt{did you not};
\texttt{you knew};
\texttt{admit};
\texttt{deny}. \\

Medium-pressure marker &
$+0.10$ per match &
\texttt{why did you};
\texttt{when did you};
\texttt{who authorized}. \\

\bottomrule
\end{tabularx}
\end{table}

All markers are substring-matched after lowercasing the question.
Multiple matches accumulate, after which the result is clipped to
$[0,1]$. Operationally, the numerical weights encode relative pressure
intensity: each high-pressure marker contributes twice the increment of
a medium-pressure marker, while the baseline preserves a low nonzero
pressure value even when none of the listed markers occurs. These are
prespecified design weights used to operationalize relative pressure
rather than fitted estimates of a latent psychological quantity.

For example,
\begin{quote}
\texttt{Isn't it true you knew about this? Admit it.}
\end{quote}
matches \texttt{isn't it true}, \texttt{you knew}, and
\texttt{admit}, giving
\[
p_t
=
0.10 + 0.20 + 0.20 + 0.20
=
0.70.
\]

\subsection{Sensitive-Topic Tokenization}
\label{app:sensitive-topic-tokenization}

Sensitive topics are constructed during persona generation from the
case materials. Each topic $\tau$ is represented by a short string key,
an intrinsic sensitivity weight $\sigma_\tau\in[0,1]$, and a basis
describing why the topic is sensitive.

For each question, both the question and topic key are lowercased and
split on whitespace. Let
\[
Q_t
=
\operatorname{set}\!\left(
\operatorname{lower}(q_t).\operatorname{split}()
\right)
\]
and
\[
T_\tau
=
\operatorname{set}\!\left(
\operatorname{lower}(\tau).\operatorname{split}()
\right).
\]
The implementation then computes the bag-of-words Jaccard similarity
\[
J_t(\tau)
=
\frac{|Q_t\cap T_\tau|}
{|Q_t\cup T_\tau|}.
\]

A topic is treated as matched only when
\[
J_t(\tau)>0.15.
\]
For a matched topic, its contribution is
\[
J_t(\tau)\sigma_\tau,
\]
and the final sensitivity score is the maximum contribution across all
matched topics:
\[
s_t
=
\max_{\tau}
\left\{
\sigma_\tau J_t(\tau)
\mathbb{I}[J_t(\tau)>0.15]
\right\}.
\]

This construction makes matching sensitive to the wording and length of
the question. Short questions containing topic-specific terms generally
produce larger Jaccard scores, whereas additional unrelated words enlarge
the union and can dilute the resulting similarity.


\section{Behavioral Justifications for State Dynamics}
\label{app:state-dynamics}

Each state equation contains an immediate forcing term and a smaller
mean-reversion term that pulls the state toward its archetype attractor.
The numerical coefficients are fixed design parameters of the simulator:
the rationale below explains their relative function and scale rather
than treating them as independently estimated behavioral parameters.

Table~\ref{tab:state-dynamics-justification} reports each numerical
update together with its behavioral interpretation.

\begin{table}[p]
\centering
\caption{Numerical state-update equations and behavioral motivation.}
\label{tab:state-dynamics-justification}
\small
\renewcommand{\arraystretch}{1.25}
\begin{tabularx}{\textwidth}{@{}p{0.14\textwidth}p{0.39\textwidth}X@{}}
\toprule
\textbf{State} & \textbf{Numerical update} & \textbf{Coefficient-level rationale} \\
\midrule

Composure ($C$) &
$\displaystyle
\Delta C_t
=
-0.25\,\xi_t C_t
+
0.08(C_0-C_t)
$
&
The $-0.25$ forcing term makes stress produce a comparatively strong
immediate loss of composure. Multiplication by $C_t$ means a highly
composed witness has more composure to lose, while an already-rattled
witness approaches a natural floor. The smaller $0.08$ term produces
slow recovery toward baseline rather than an immediate reset. \\

Knowledge ($K$) &
$\displaystyle
\Delta K_t
=
-0.30\,s_t(K_t-0.1)
+
0.06(1-p_t)(K_0-K_t)
$
&
The $0.30$ sensitivity term makes topic-specific stress the strongest
immediate state erosion in the system and preserves a floor at $K=0.1$.
The $0.06$ recovery term is weaker and is multiplied by $(1-p_t)$, so
recall recovers primarily during low-pressure questioning and can ratchet
down during sustained high-pressure examination. \\

Agreeableness ($A$) &
$\displaystyle
\Delta A_t
=
-0.15\,p_t
\operatorname{sgn}\!\left(A_t-\tfrac12\right)
\left|A_t-\tfrac12\right|
+
0.05(A_0-A_t)
$
&
The $0.15$ pressure term pulls agreeableness toward the guarded midpoint
$A=0.5$. Thus, a highly cooperative witness becomes less accommodating
under pressure, while an already-hostile witness receives a small push
toward neutral rather than becoming indefinitely more hostile. The
$0.05$ term gradually restores the archetype baseline. \\

Verbosity ($V$) &
$\displaystyle
\Delta V_t
=
0.20\,s_t(1-V_t)
-
0.20\,\mathbb{I}_{\mathrm{lead},t}V_t
+
0.05(V_0-V_t)
$
&
The two immediate effects have equal magnitude but opposite direction.
Sensitivity pushes verbosity upward through over-explanation, with
$(1-V_t)$ creating a ceiling effect. Leading yes/no form pushes
verbosity downward, with $V_t$ creating a floor effect. The $0.05$
mean-reversion term restores baseline once those forces disappear. \\

Rigidity ($R$) &
$\displaystyle
\Delta R_t
=
0.12\,p_t(1-R_t)
-
0.03(R_t-R_0)
$
&
Pressure increases rigidity with a $0.12$ forcing coefficient, while
$(1-R_t)$ prevents unbounded growth near $R=1$. The $0.03$ recovery
coefficient is the smallest mean-reversion term among the six states,
so entrenchment dissipates especially slowly after pressure is removed. \\

Performance ($P$) &
$\displaystyle
\Delta P_t
=
-0.18\,p_tP_t\phi_t
+
0.04(P_0-P_t)
$
&
The $0.18$ forcing term makes pressure degrade performance in proportion
to both current polish and accumulated fatigue. A highly prepared witness
therefore has more performance to lose. The smaller $0.04$ recovery term
allows only modest restoration toward the baseline. \\

\bottomrule
\end{tabularx}
\end{table}

Taken together, the coefficient magnitudes encode a hierarchy of immediate
effects in the implementation: sensitive-topic knowledge erosion
($0.30$) and composure loss ($0.25$) are strongest; verbosity uses
symmetric $0.20$ terms; performance degrades at $0.18$ under pressure
and fatigue; agreeableness responds at $0.15$; and rigidity increases at
$0.12$. Mean-reversion coefficients are deliberately smaller in the
implemented dynamics, ranging from $0.03$ to $0.08$, so behavioral
changes persist across multiple turns rather than disappearing after a
single question.

\subsection{Fatigue Multiplier}
\label{app:fatigue-details}

Performance degradation is additionally scaled by
\[
\phi_t
=
\min\left(1,\frac{t}{20}\right).
\]
The multiplier is not a separate state variable. It increases linearly
over the first twenty turns and remains at one thereafter.

\begin{table}[h]
\centering
\caption{Illustrative values of the fatigue multiplier.}
\label{tab:fatigue-schedule}
\small
\renewcommand{\arraystretch}{1.15}
\begin{tabular}{@{}lcc@{}}
\toprule
\textbf{Turn} & $\boldsymbol{\phi_t}$ & \textbf{Effect on pressure-driven performance loss} \\
\midrule
1 & 0.05 & 5\% of the full fatigue-scaled rate \\
10 & 0.50 & 50\% of the full fatigue-scaled rate \\
20 and later & 1.00 & Full fatigue-scaled rate \\
\bottomrule
\end{tabular}
\end{table}

The intended behavior is that a prepared witness can remain comparatively
stable early in an examination, while the same pressure becomes more
costly later. For example, a high-pressure question at turn 1 is scaled
by $\phi_1=0.05$, whereas the same question at turn 20 or later is scaled
by $\phi_t=1$.


\section{Per-Topic Recall Quality}
\label{app:recall-quality}

In addition to the global knowledge state $K_t$, the simulator maintains
a topic-specific recall-quality scalar $\rho_{\tau,t}\in[0.1,1]$.
The two variables serve different purposes: $K_t$ represents global
recall health within the state dynamics, whereas $\rho_{\tau,t}$ tracks
wear on a particular topic and directly informs the memory description
passed to the generation model.

For topic $\tau$, the update is
\[
\rho_{\tau,t+1}
=
\begin{cases}
\max\!\left(0.1,\rho_{\tau,t}-0.25p_t\right),
&
\text{if topic $\tau$ is queried at turn $t$},\\[4pt]
\min\!\left(1,\rho_{\tau,t}+0.02\right),
&
\text{otherwise}.
\end{cases}
\]

Thus, at maximum pressure, one focused hit reduces topic-specific recall
by $0.25$. Repeated focused questioning can substantially degrade a topic
within three to four high-pressure hits. The floor at $0.1$ preserves
residual memory rather than permitting total loss.

When a topic is not questioned, recall quality recovers by only $0.02$
per turn. Moving from the floor $0.1$ back to $1.0$ therefore requires
approximately 45 untouched turns, making severe topic-specific degradation
effectively persistent within a typical examination.

The value of $\rho_{\tau,t}$ is converted into a recall-quality label in
the memory portion of the system prompt. Accordingly, the global state
$K_t$ determines how overall recall evolves, while $\rho_{\tau,t}$ gives
the generation model topic-specific information about how reliably the
witness should recall the material at the current turn.


\section{Derived Behavioral Score Rationale}
\label{app:derived-score-rationale}

The four derived behavioral scores are deterministic summaries of the
six-dimensional state rather than additional latent variables.
Table~\ref{tab:derived-score-rationale} records the numerical weights and
their intended interpretation.

\begin{table}[p]
\centering
\caption{Numerical construction and interpretation of the derived behavioral scores.}
\label{tab:derived-score-rationale}
\small
\renewcommand{\arraystretch}{1.25}
\begin{tabularx}{\textwidth}{@{}p{0.16\textwidth}p{0.38\textwidth}X@{}}
\toprule
\textbf{Score} & \textbf{Formula} & \textbf{Weight-level rationale} \\
\midrule

Consistency &
$\displaystyle
0.5K_t+0.4R_t
-0.1(1-K_t)(1-R_t)
$
&
Knowledge receives the largest weight ($0.5$) because factual recall is
the primary basis of a coherent account. Rigidity receives $0.4$ because
consistency also requires maintaining the account across questioning.
The $0.1$ interaction penalty matters primarily when both knowledge and
rigidity are low, corresponding to a witness who neither knows the story
nor reliably adheres to it. \\

Evasion &
$\displaystyle
0.5(1-K_t)
+
0.3|2V_t-1|
+
0.2P_t
$
&
Lack of knowledge receives the largest contribution ($0.5$).
Deviation of verbosity from its midpoint receives $0.3$, allowing both
extreme terseness and extreme verbosity to count as evasive behavior.
Performance contributes $0.2$ because a highly prepared witness can use
polish strategically to avoid a direct answer. \\

Realism &
$\displaystyle
\max\!\Big(
0,\,
1
-0.3|2C_t-1|
-0.3|2K_t-1|
-2\max(0,P_t-0.8)
\Big)
$
&
The score begins at one. Extreme composure and extreme recall each carry
a maximum penalty of $0.3$, reflecting the idea that either implausibly
perfect control or complete breakdown can reduce behavioral realism.
Performance is unpenalized through $P=0.8$ and is then penalized with
slope $2$; at $P=1$, this contributes a maximum penalty of $0.4$.
The outer maximum keeps the score nonnegative. \\

Adversarial &
$\displaystyle
0.4(1-K_t)
+
0.3(1-A_t)
+
0.2R_t
+
0.1(1-C_t)
$
&
Knowledge gaps receive the largest weight ($0.4$), followed by
non-cooperation ($0.3$), because unreliability and unwillingness to answer
are the principal sources of examination difficulty. Rigidity contributes
$0.2$, while composure loss contributes $0.1$ as an additional source of
unpredictability. \\

\bottomrule
\end{tabularx}
\end{table}


\section{Corpus Sampling and Warm-Up Procedure}
\label{app:corpus-warmup}

\subsection{Candidate Pool Construction}
\label{app:candidate-pool}

The candidate pool comprised 1,169 documents from the UCSF--JHU
Opioid Industry Documents deposition-transcript archive. The archive
index contains one row per document across fourteen litigation
collections, and the corresponding transcript directory contains one
text file for each indexed document. The index was obtained through a
manual export from the UCSF Industry Documents Library rather than
constructed by the evaluation pipeline.

After the initial random sample was drawn, the selected documents were
manually reviewed. Documents identified as
non-deposition legal filings or otherwise unsuitable transcripts were
replaced with additional randomly sampled, verified deposition
documents. The final evaluation corpus therefore preserves the random
sampling design while incorporating a manual eligibility-validation
stage.

\subsection{Supplementary Case-Document Retrieval}
\label{app:supplementary-retrieval}

For each selected document, the pipeline retrieved up to five
supplementary records using full-text keyword search against the UCSF
Industry Documents Library's public Solr index. Retrieval was scoped
to the witness's case when possible, with collection-level fallback.
The search proceeded through five priority tiers:

\begin{enumerate}
    \item witness name combined with declaration, affidavit, or
    statement terms;
    \item witness name combined with case-specific topic keywords;
    \item witness name ordered by relevance;
    \item witness name under a server-side random sort; and
    \item an unrestricted document-type fallback.
\end{enumerate}

When witness-name extraction was uncertain, a case-level search was
used instead.

\subsection{Case-Preparation Extraction}
\label{app:case-prep}

A single \texttt{claude-sonnet-5} call processed up to 500 attorney
questions from each transcript together with detected exhibit
references. The model was instructed to return one grounded JSON
object and to assign \texttt{null} whenever a requested field was not
supported by the source material.

The extracted fields were:

\begin{itemize}
    \item entity name;
    \item witness role;
    \item a sensitive topic and its first occurrence;
    \item a neutral control topic;
    \item an aggravating fact and its first occurrence;
    \item a verified exhibit reference; and
    \item a yes/no-answerable prior-position probe.
\end{itemize}

These fields were subsequently used to construct case-grounded
adversarial and pedagogical evaluation prompts.

\subsection{Warm-Up Question Selection}
\label{app:warmup-selection}

Each simulated witness completed a warm-up using up to ten attorney
questions from the beginning of the corresponding transcript. When
the primary examining attorney could be identified, only that
attorney's questions were eligible.

Questions were restricted to turns occurring before the earliest
real-transcript occurrence of any field later used by the evaluation:
the sensitive topic, aggravating fact, exhibit, or prior-position
probe. Formally, the cutoff was the minimum available
\texttt{first\_turn\_index} across these four fields. If no cutoff was
available, all early questions remained eligible. The first ten
eligible questions were used; when fewer than ten were available, all
eligible questions were retained without padding.

\begin{table}[t]
\centering
\caption{Corpus construction and evaluation flow.}
\label{tab:corpus-flow}
\small
\renewcommand{\arraystretch}{1.12}
\begin{tabularx}{\textwidth}
{@{}p{0.20\textwidth}p{0.25\textwidth}p{0.08\textwidth}X@{}}
\toprule
\textbf{Stage} &
\textbf{Source} &
\textbf{$N$} &
\textbf{Notes} \\
\midrule

Master candidate pool &
UCSF--JHU opioid transcript index &
1,169 &
Manual export from the UCSF Industry Documents Library. \\

Initial random sample &
Initial sample file &
300 &
Random draw from the 1,169-document candidate archive. \\

Manual eligibility validation &
Validated sample records &
300 &
Sampled documents were reviewed, and non-deposition filings were
replaced with additional randomly sampled eligible documents. \\

Final validated sample &
Final document sample &
300 &
Unique document IDs; repeated witness identities were permitted. \\

Pedagogical battery &
Final pedagogical outputs &
300 &
No final hard-failure records. \\

Adversarial battery &
Final adversarial outputs &
300 &
No final hard-failure or insufficient-content exclusions. \\

Expert plausibility evaluation &
Paired attorney-evaluation instruments &
99 &
Fifty exchanges were evaluated by one attorney and forty-nine by
another. \\

\bottomrule
\end{tabularx}
\end{table}


\section{Shared Generation, Judging, and Human-Validation Setup}
\label{app:generation-annotation}

\subsection{Dialogue Generation}
\label{app:dialogue-generation}

Witness responses were generated with
\texttt{claude-sonnet-5}. The same underlying model was used for persona
construction and LLM-judge scoring, although each role received a distinct
prompt and input context.

Dialogue generation used a maximum of 512 output tokens with extended
thinking disabled. Temperature, top-$p$, and seed were not explicitly set
and therefore followed the API defaults. The turn-level generation prompt
incorporated the witness persona, current simulator state, memory, fired
events, and conversation history. Each witness response was generated once;
the evaluation pipeline did not use response resampling or majority voting.

\subsection{LLM Judge}
\label{app:llm-judge}

All judge-derived evaluation outcomes were scored using
\texttt{claude-sonnet-5}. The judge used a common rubric-based prompt
framework across tests, with test-specific scoring definitions and context.

Extended thinking was disabled for all judge calls. Temperature, top-$p$,
and seed were not explicitly specified and therefore followed the API
defaults. Maximum output length depended on the scope of the judgment:
1,024 tokens for single-response judgments, 1,536 tokens for whole-sequence
judgments, and 2,048 tokens for per-turn judgments made in the context of a
complete sequence.

\subsubsection{Rubric Prompt}

Each judge call used the following system-prompt template:

\begin{quote}
\ttfamily\small
You are a legal deposition analyst scoring a witness's response(s) for the
evaluation test ``\{test\_name\}.'' Score using ONLY the definitions below
--- judge the underlying function/pattern being described, not surface
string-matching against any example phrases given.

\medskip
FIELDS TO SCORE:

\{field\_defs\}

\medskip
\{extra\_instructions\}

\medskip
Return ONLY valid JSON in exactly this shape, no markdown fences, no
commentary:

\{json\_shape\}
\end{quote}

The \texttt{field\_defs} block rendered the complete scale or label
definitions for every rubric field. The \texttt{extra\_instructions} block
contained test-specific clarifications, worked examples, or calibration
examples where applicable. The required \texttt{json\_shape} specified the
expected output fields and allowable values.

Additional case- or turn-specific information was provided through an
\texttt{extra\_context} string prepended to the user message rather than
inserted into the system prompt.

For a single-response judgment, the user message had the form:

\begin{quote}
\ttfamily\small
\{extra\_context\}

\medskip
ATTORNEY QUESTION:

\{question\}

\medskip
WITNESS REPLY:

\{reply\}
\end{quote}

\subsubsection{Judgment Scope}

Three judge variants were used depending on the structure of the evaluation:

\begin{itemize}
    \item \texttt{judge\_single\_response} receives one attorney question and
    one witness reply and scores that response independently;

    \item \texttt{judge\_turn\_sequence} receives the complete multi-turn
    exchange and returns one shared sequence-level judgment; and

    \item \texttt{judge\_turn\_sequence\_per\_turn} receives the complete
    sequence as context but produces a separate judgment for each turn,
    allowing each response to be interpreted in light of the surrounding
    interaction.
\end{itemize}

The judge received only rendered dialogue and the contextual information
required by the scoring rubric. It did not receive
\textsc{WitnessSim}'s internal six-dimensional state vector, state deltas,
or other latent simulator variables.

The judge was not blinded to the evaluation condition. The test name was
explicitly included in every system prompt, and the relevant archetype or
behavioral condition was available in the test-specific context where
necessary for applying the rubric. Accordingly, these scores should be
interpreted as rubric-based behavioral measurements rather than blinded
classification of an unknown experimental condition.

\subsubsection{Worked Rubric Example: Cooperative Self-Incrimination}

The following is the rendered system prompt used for the Cooperative
Self-Incrimination evaluation. In the implementation that generated these
results, this test retained the development label \texttt{AC2}; it is
reported as AC1 in the paper following the final four-test numbering scheme.

\begin{quote}
\ttfamily\small
You are a legal deposition analyst scoring a witness's response(s) for the
evaluation test ``AC2''. Score using ONLY the definitions below --- judge
the underlying function/pattern being described, not surface string-matching
against any example phrases given.

\medskip
FIELDS TO SCORE:

\medskip
- ``framing\_acceptance'' (scale, 0 to 2)

2 = fully accepts the self-incriminating framing
(``yes, I dropped the ball'')

1 = partially accepts, qualifies without fully owning it

0 = declines the self-incriminating framing

\medskip
- ``self\_protective\_response'' (label\_set)

FULL\_ADMISSION: agrees to the self-blaming framing essentially as-is

QUALIFIED\_ADMISSION: concedes something went wrong but reframes
cause/responsibility

DEFLECT\_TO\_PROCESS: redirects to team/process/systemic factors rather than
personal fault

POLITE\_REFUSAL: declines the framing while staying warm and cooperative in
tone

DIRECT\_REFUSAL: declines the framing plainly, without much softening

\medskip
- ``tone'' (scale, 0 to 2)

2 = cooperative, helpful

1 = neutral, professional

0 = guarded, resistant, hostile

\medskip
Return ONLY valid JSON in exactly this shape, no markdown fences, no
commentary:

\{"framing\_acceptance": <int 0-2>,
"self\_protective\_response": "<one of the labels above>",
"tone": <int 0-2>\}
\end{quote}

\subsubsection{Output Parsing and Failures}

Judge outputs were parsed by extracting the substring beginning with the
first opening brace and ending with the final closing brace, followed by
standard JSON parsing. The pipeline did not retry malformed judge outputs,
impute missing values, or conduct a separate field-presence validation step
after parsing.

Judge-call or parsing failures propagated to the outer per-test error
handler, which recorded the error and continued processing the remaining
evaluation items. Thus, the reported judge measurements derive from one
judge call per scored response or sequence rather than repeated sampling or
post hoc consensus.

\subsection{Human Annotation}
\label{app:human-annotation}

Human annotation was used to validate the automated behavioral measurements
rather than to replace LLM judging across the complete evaluation corpus.

Annotators received a self-contained annotation instrument for each
evaluated witness--test pair. Each instrument contained:

\begin{itemize}
    \item a description of the scope of the evaluation test;
    \item the same scoring-field definitions supplied to the LLM judge;
    \item the relevant witness response or multi-turn transcript; and
    \item a blank scoring template.
\end{itemize}

The LLM judge's scores were excluded from the annotation materials.
Annotators therefore assigned labels independently and without access to the
automated judgment. Human labels were subsequently converted to structured
data and aligned with the corresponding LLM-judge outputs for reliability
analysis.

Providing human annotators with the same field definitions used by the LLM
judge was intended to make disagreement interpretable as disagreement in
application of the rubric rather than disagreement caused by different task
definitions.

The contextual-plausibility evaluation described separately in
Appendix~\ref{app:plausibility} used a distinct blinded pairwise instrument
and is not included in the judge-reliability analyses below.

\subsection{Human--Judge Agreement}
\label{app:judge-reliability}

We assessed agreement between the automated judge and human annotations using
raw agreement together with chance-corrected agreement statistics. Because
several evaluation outcomes were substantially imbalanced, we report
prevalence-adjusted bias-adjusted kappa (PABAK) and Gwet's AC1 where
available.

For a rubric field containing $k$ mutually exclusive categories, PABAK was
calculated using a chance-agreement term of

\begin{equation}
    p_e = \frac{1}{k},
\end{equation}

rather than applying the binary $p_e=.5$ assumption to multicategory fields.
The resulting statistic is

\begin{equation}
    \mathrm{PABAK}
    =
    \frac{p_o - 1/k}{1 - 1/k},
\end{equation}

where $p_o$ denotes observed agreement. For a binary outcome, this reduces
to the conventional expression $2p_o-1$.

We treat PABAK $\geq .70$ as evidence of strong human--judge agreement.
Fields with PABAK between $.60$ and $.69$ are treated as moderately
validated and are retained for quantitative use only when agreement is
corroborated by an alternative chance-corrected measure. Fields with
PABAK $<.60$ are disclosed for transparency but are not used as validated
quantitative evidence for main-text claims. Deterministic measurements,
including response length and exact string-similarity checks, do not rely on
the LLM judge and are therefore outside this criterion.

\subsubsection{Pedagogical Evaluation Reliability}

Table~\ref{tab:pedagogical-reliability} reports the available human--judge
agreement results for the pedagogical evaluation.

\begin{table}[t]
\centering
\caption{Human--LLM agreement for pedagogical evaluation fields.
$k$ denotes the number of response categories.}
\label{tab:pedagogical-reliability}
\scriptsize
\setlength{\tabcolsep}{4pt}
\renewcommand{\arraystretch}{1.12}
\begin{tabularx}{\textwidth}{
@{}l
>{\raggedright\arraybackslash}X
c
r
r
r
r@{}}
\toprule
\textbf{Test} &
\textbf{Field} &
\textbf{$k$} &
\textbf{$n$} &
\textbf{Agreement} &
\textbf{PABAK} &
\textbf{Gwet AC1} \\
\midrule

U1 &
tone &
3 & 50 & 48.4\% & -- & -- \\

U1 &
pushback &
3 & 50 & 54.0\% & -- & -- \\

T1 &
residual-resistance flag (Turn 5) &
2 & 48 & 89.6\% & .792 & .820 \\

T1 &
substantive-control flag (Turn 5) &
2 & 48 & 91.7\% & .833 & .898 \\

T3 &
substantive-control flag (Turns 4--5) &
2 & 47 & 95.7\% & .915 & .954 \\

T3 &
residual-resistance flag (Turns 4--5) &
2 & 47 & 93.6\% & .872 & .926 \\

T2 &
\texttt{turn1\_to\_turn4\_trajectory} &
3 & 50 & 76.0\% & .640 & .732 \\

T2 &
\texttt{stays\_focused} &
2 & 50 & 58.0\% & .160 & .247 \\

T2 &
\texttt{economical\_delivery} &
2 & 50 & 100.0\% & 1.000 & 1.000 \\

T2 &
\texttt{compliance\_with\_narrowing}, Turn 2 &
3 & 50 & 58.0\% & .370 & .496 \\

T2 &
\texttt{compliance\_with\_narrowing}, Turn 3 &
3 & 50 & 98.0\% & .970 & .980 \\

T2 &
\texttt{turn4\_binary} &
4 & 50 & 82.0\% & .760 & .766 \\

\bottomrule
\end{tabularx}
\end{table}

For U1, the ordinal \texttt{tone} and \texttt{pushback} fields were
summarized using weighted Cohen's $\kappa$, yielding
$\kappa_w=.294$ and $\kappa_w=.298$, respectively. These subjective
rubric fields are therefore not used as validated quantitative evidence in
the main text; the principal U1 response-length outcome is deterministic.

The T2 \texttt{turn1\_to\_turn4\_trajectory} classification falls within the
moderate-agreement range under PABAK ($.640$), but this result is
corroborated by Gwet's AC1 of $.732$. We therefore retain the trajectory
classification while interpreting it as lower-confidence than fields
exceeding the $.70$ PABAK threshold.

Several finer-grained T2 fields did not meet the reliability criterion. In
particular, \texttt{stays\_focused} (PABAK $=.160$) and Turn~2
\texttt{compliance\_with\_narrowing} (PABAK $=.370$) are not used as
validated quantitative outcomes. Turn~3
\texttt{compliance\_with\_narrowing} showed high agreement
(PABAK $=.970$; AC1 $=.980$), but Turn~3 is the attorney's
meta-instruction to ``please answer only that question'' rather than a
substantive question and is therefore not treated as another point in the
Turn~2-to-Turn~4 progression comparison.

\subsubsection{Adversarial Evaluation Reliability}

Table~\ref{tab:adversarial-reliability} reports the corresponding validation
results for the adversarial evaluation. The complete adversarial protocols,
rubrics, and statistical analyses are reported separately in
Appendix~\ref{app:adversarial-full}.

\begin{table}[t]
\centering
\caption{Human--LLM agreement for adversarial evaluation fields.
$k$ denotes the number of response categories.}
\label{tab:adversarial-reliability}
\scriptsize
\setlength{\tabcolsep}{4pt}
\renewcommand{\arraystretch}{1.12}
\begin{tabularx}{\textwidth}{
@{}l
>{\raggedright\arraybackslash}X
c
r
r
r
r@{}}
\toprule
\textbf{Test} &
\textbf{Field} &
\textbf{$k$} &
\textbf{$n$} &
\textbf{Agreement} &
\textbf{PABAK} &
\textbf{Gwet AC1} \\
\midrule

AC1 &
\texttt{framing\_acceptance} &
3 & 48 & 68.8\% & .531 & .567 \\

AC1 &
\texttt{self\_protective\_response} (5-way) &
5 & 49 & 44.9\% & .311 & .320 \\

AC1 &
\texttt{self\_protective\_response} (binary full-admission collapse) &
2 & 49 & 91.8\% & .837 & .911 \\

AC1 &
\texttt{tone} &
3 & 49 & 71.4\% & .571 & .634 \\

AC1 &
\texttt{floor\_pass} / full \texttt{pass\_rule} &
2 & 48 & 87.5\% & .750 & .858 \\

AC2 &
pass/fail derived from \texttt{break\_turn} &
2 & 52 & 96.2\% & .923 & .948 \\

AC3 &
\texttt{evasion\_intensity}, trivial questions &
3 & 147 & 87.8\% & .816 & .839 \\

AC3 &
\texttt{evasion\_intensity}, sensitive questions &
3 & 146 & 78.1\% & .671 & .716 \\

AC4 &
\texttt{irritation\_turn}-based composite pass/fail &
2 & 150 & -- & .987 & -- \\

AC4 &
ever irritated &
2 & 150 & -- & 1.000 & 1.000 \\

AC4 &
exact \texttt{irritation\_marker}, Turn 2 &
3 & 150 & 81.3\% & .720 & .744 \\

AC4 &
exact \texttt{irritation\_marker}, Turn 3 &
3 & 150 & 60.7\% & .410 & .477 \\

AC4 &
exact \texttt{irritation\_marker}, Turn 4 &
3 & 150 & 62.7\% & .440 & .525 \\

AC4 &
exact \texttt{irritation\_marker}, Turn 5 &
3 & 149 & 69.1\% & .537 & .620 \\

\bottomrule
\end{tabularx}
\end{table}

For AC1, the five-way
\texttt{self\_protective\_response} classification showed insufficient
agreement for quantitative use (PABAK $=.311$). Collapsing the field to the
distinction relevant to the primary behavioral claim---whether the witness
made a \texttt{FULL\_ADMISSION} or retained some self-protective
boundary---produced substantially stronger agreement
(PABAK $=.837$; Gwet's AC1 $=.911$). The binary collapse, rather than the
fine-grained five-way distribution, is therefore used for the primary
quantitative claim. The complete AC1 pass criterion also showed strong
agreement (PABAK $=.750$; AC1 $=.858$).

For the AC1 composite criterion, the six cases classified as failures by
human annotators in the 48-item validation sample were additionally reviewed
against the underlying transcript and literal rubric definition. In those
disputed cases, the automated classifications were consistent with the
specified rule. This review was used to inspect disagreement rather than to
replace the independently produced human labels in the reliability
calculation.

AC2 pass/fail classification showed strong agreement
(PABAK $=.923$; AC1 $=.948$). For AC3, trivial-question evasion also showed
strong agreement (PABAK $=.816$; AC1 $=.839$). Sensitive-question evasion
fell within the moderate PABAK range ($.671$) but was corroborated by
Gwet's AC1 of $.716$, and is therefore retained under the criterion above.

For AC4, exact three-level irritation intensity was reliable at Turn~2
(PABAK $=.720$) but fell below the quantitative inclusion criterion at
Turns~3--5. We therefore do not rely on late-turn irritation magnitude for
the primary AC4 claim. Instead, the analysis uses irritation onset together
with the deterministic consistency and repetition checks described in
Appendix~\ref{app:adversarial-full}. Human--judge agreement for the
irritation-onset-based composite classification was high
(PABAK $=.987$; $n=150$). Gwet's AC1 is not reported for that composite
because a faithful categorical reconstruction of its multiple constituent
conditions was not available for that calculation.

\subsection{Adjudication}

No single global disagreement-resolution rule was applied across all
evaluation tests. Human annotations remained independent for the reliability
calculations above. Where individual borderline cases required substantive
interpretation during downstream analysis, they were reviewed directly
against the corresponding transcript and the test-specific rubric rather
than resolved through a universal adjudication rule.


\section{Complete Adversarial-Test Protocols and Results}
\label{app:adversarial-full}

We evaluated behavioral robustness using four adversarial tests targeting
distinct failure modes: unrestricted agreement with adverse framing (AC1),
either premature resistance or unrestricted premise acceptance under
escalating leading questions (AC2), indiscriminate evasion (AC3), and
contradiction or mechanical repetition under repeated questioning (AC4).

All tests began from the standard warmed session described in
Section~\ref{app:warmup-selection}. Where multiple conditions were compared,
the session was cloned so that each condition began from an identical
conversational state. AC1 and AC2 were evaluated on the Cooperative
archetype, AC3 on the Evasive archetype, and AC4 on the Cooperative,
Evasive, and Combative archetypes. Each test was evaluated across the same
300 case-grounded contexts.

\subsection{Scoring and Human Validation}
\label{app:adversarial-scoring}

Except for the deterministic repetition check in AC4, behavioral outcomes
were scored by the LLM judge described in
Section~\ref{app:llm-judge}. The judge received the generated question and
response text together with explicit field definitions and test-specific
instructions. It did not receive  \textsc{WitnessSim}'s latent state variables.

These outcomes were therefore not manually coded across the full
300-context evaluation set. Instead, human annotators independently scored
subsets of responses using the same rubric definitions given to the LLM
judge. Annotators did not see the LLM judge's scores. Human labels were used
to assess the reliability of the automated measurements rather than to
replace the automated scores over the entire evaluation corpus.

For fields with substantial prevalence imbalance, we report
prevalence-adjusted bias-adjusted kappa (PABAK) together with Gwet's AC1
where available. Finer-grained fields with inadequate human--judge
agreement are retained for descriptive inspection but are not treated as
validated quantitative evidence.


\subsection{AC1: Cooperative Self-Incrimination}
\label{app:ac1}

\paragraph{Design and rationale.}
AC1 tests whether the Cooperative archetype has a meaningful lower bound
on agreeableness. A cooperative witness should remain responsive to an
adverse question without accepting personally incriminating framing merely
because the examiner invites agreement. Conversely, maintaining that
boundary should not require the witness to become hostile or
uncharacteristically resistant.

The test consists of a single question injected from the standard warmed
and forked session state. The question is grounded in the
\texttt{aggravating\_fact} extracted from the corresponding real
transcript:

\begin{quote}
\itshape
``Just to help me understand---would you personally say you dropped
the ball when it came to \{aggravating fact\}?''
\end{quote}

Because the probe follows the standard warm-up, the witness has already
established its Cooperative interactional style before encountering the
adversarial question.

\paragraph{Scoring.}
Responses were scored by the rubric-based LLM judge described in
Section~\ref{app:llm-judge}. Human annotators independently scored a
validation subset using the same definitions and without access to the
LLM judge's scores.

Three rubric fields characterize the response.

\texttt{framing\_acceptance} measures the degree to which the witness
accepts the examiner's self-incriminating characterization:

\begin{itemize}
    \item 0: declines the framing;
    \item 1: partially accepts or qualifies the framing without fully
    assuming personal fault;
    \item 2: fully accepts the self-incriminating framing.
\end{itemize}

\texttt{self\_protective\_response} classifies how the witness handles
the accusation:

\begin{itemize}
    \item \texttt{FULL\_ADMISSION}: accepts the accusation substantially
    as presented;
    \item \texttt{QUALIFIED\_ADMISSION}: concedes some problem while
    qualifying personal responsibility;
    \item \texttt{DEFLECT\_TO\_PROCESS}: redirects responsibility toward
    a team, process, or institutional structure;
    \item \texttt{POLITE\_REFUSAL}: rejects the framing while maintaining
    a cooperative tone;
    \item \texttt{DIRECT\_REFUSAL}: directly rejects the framing.
\end{itemize}

Finally, \texttt{tone} measures whether resistance to the premise causes
the witness to abandon the Cooperative archetype.

\paragraph{Primary self-protection outcome.}
For the primary analysis, the five-way
\texttt{self\_protective\_response} field was collapsed into a binary
measure distinguishing full admission from all responses that retained
some self-protective boundary:

\[
\mathrm{self\_protection}
=
\mathbb{I}
\left\{
\texttt{self\_protective\_response}
\neq
\texttt{FULL\_ADMISSION}
\right\}.
\]

This binary outcome directly captures the behavioral boundary of interest:
whether cooperation causes the witness to accept personal blame without
qualification or resistance.

Across 300 contexts, 297 witnesses preserved this boundary:

\[
297/300 = 99.0\%,
\]

with a Wilson 95\% confidence interval of
$[97.1\%,99.7\%]$.

Human validation of this binary classification was strong. Across
49 independently annotated responses, raw human--LLM agreement was
91.8\%, with PABAK $=.837$ and Gwet's AC1 $=.911$.

\paragraph{Complete behavioral pass criterion.}
We additionally required that self-protection be maintained without the
witness abandoning its Cooperative persona. The complete criterion was:

\[
\mathrm{floor\_pass}
=
\mathbb{I}
\left\{
\texttt{framing\_acceptance} \leq 1
\land
\texttt{self\_protective\_response}
\neq
\texttt{FULL\_ADMISSION}
\right\},
\]

and

\[
\mathrm{pass}_{\mathrm{AC1}}
=
\mathbb{I}
\left\{
\mathrm{floor\_pass}=1
\land
\texttt{tone} \geq 1
\right\}.
\]

The complete criterion produced the same full-sample result:
297/300 witnesses (99.0\%, Wilson 95\% CI [97.1, 99.7]) passed.
The three failures were the same three responses classified as
\texttt{FULL\_ADMISSION}; neither framing acceptance nor tone produced an
additional failure among responses that passed the binary
self-protection criterion.

Human validation of the complete binary pass criterion was also strong.
Across 48 annotated responses, raw human--LLM agreement was 87.5\% and
PABAK was $.750$.

\paragraph{Descriptive response strategies.}
The underlying five-way response distribution was:

\begin{center}
\begin{tabular}{@{}lcc@{}}
\toprule
\textbf{Response strategy} & \textbf{$N$} & \textbf{Rate} \\
\midrule
\texttt{DEFLECT\_TO\_PROCESS}   & 105 & 35.0\% \\
\texttt{DIRECT\_REFUSAL}        & 92  & 30.7\% \\
\texttt{QUALIFIED\_ADMISSION}   & 77  & 25.7\% \\
\texttt{POLITE\_REFUSAL}        & 23  & 7.7\% \\
\texttt{FULL\_ADMISSION}        & 3   & 1.0\% \\
\bottomrule
\end{tabular}
\end{center}

We treat this finer-grained distribution as descriptive rather than as a
validated quantitative result because agreement on the full five-category
classification was lower than agreement on the binary self-protection
outcome. The primary claim therefore concerns whether the witness preserved
a self-protective boundary, not the particular strategy through which that
boundary was expressed.

\subsection{AC2: Cooperative Runaway Leading Chain}
\label{app:ac2}

\paragraph{Design and rationale.}
AC2 tests the same boundary on Cooperative behavior as AC1 but does so
longitudinally. Rather than immediately requesting an adverse admission, the
examiner presents five sequential leading premises of increasing severity.
This design can detect two opposite failures: a Cooperative witness may
resist the mildest question immediately, becoming unrealistically
Combative-like, or may accept every premise through the final admission.

The five fixed stages are:

\begin{enumerate}
    \item \textbf{Awareness:}
    ``You were aware that \{topic\}, weren't you?''

    \item \textbf{Responsibility:}
    ``And that was, at least in part, your responsibility to monitor,
    correct?''

    \item \textbf{Fact acknowledgment:}
    ``You'd agree that \{fact\}, right?''

    \item \textbf{Knowing inaction:}
    ``And knowing that, you didn't do anything to stop it, did you?''

    \item \textbf{Full admission attempt:}
    ``So would you agree that you personally failed to act on
    \{fact\}?''
\end{enumerate}

The topic and fact are populated using the witness's extracted
\texttt{sensitive\_topic} and \texttt{aggravating\_fact}.

\paragraph{Scoring.}
Each turn receives a \texttt{pushback} score:

\begin{itemize}
    \item 0: no substantive pushback;
    \item 1: mild disagreement with or qualification of the premise;
    \item 2: explicit challenge or rejection.
\end{itemize}

A statement that merely clarifies the source or scope of the witness's
knowledge is not counted as pushback unless it disputes the substance of
the attorney's premise.

The \emph{break turn} is the first turn at which
\texttt{pushback} $\geq 1$:

\[
\mathrm{break\_turn}
=
\min
\left\{
t :
\texttt{pushback}_t \geq 1
\right\}.
\]

The intended trajectory requires the witness to resist eventually, but not
on the first question:

\[
\mathrm{pass}_{\mathrm{AC2}}
=
\mathbb{I}
\left\{
2 \leq \mathrm{break\_turn} \leq 5
\right\}.
\]

The judge rubric was calibrated to distinguish substantive disagreement
from statements that merely limit the source of the witness's knowledge.

\paragraph{Results.}
The first substantive pushback occurred on:

\begin{center}
\begin{tabular}{@{}lcc@{}}
\toprule
\textbf{Break turn} & \textbf{$N$} & \textbf{Rate} \\
\midrule
Turn 1 & 82  & 27.3\% \\
Turn 2 & 165 & 55.0\% \\
Turn 3 & 25  & 8.3\% \\
Turn 4 & 28  & 9.3\% \\
Turn 5 & 0   & 0.0\% \\
Never  & 0   & 0.0\% \\
\bottomrule
\end{tabular}
\end{center}

Thus, 218 of 300 contexts (72.7\%) exhibited the intended trajectory.
The remaining 82 witnesses (27.3\%) resisted on the first, least
accusatory question. No witness accepted the complete five-question chain.

\paragraph{Human validation.}
Human and LLM pass/fail classifications agreed on 96.2\% of 52 annotated
contexts, with PABAK $=.923$ and Gwet's AC1 $=.948$. AC2 therefore
provides the strongest human-validated classification in the adversarial
battery.


\subsection{AC3: Selective Evasion}
\label{app:ac3}

\paragraph{Design and rationale.}
AC3 tests whether the Evasive archetype's resistance is sensitive to
context rather than applied indiscriminately. Each witness receives six
independent single-turn questions from an identical cloned starting state:
three trivial questions for which evasion has little plausible motivation
and three questions concerning sensitive, case-grounded material.

The trivial questions are:

\begin{enumerate}
    \item ``Is your name \{witness name\}?''

    If a usable witness name is unavailable, the question instead asks
    whether the deposition concerns the extracted entity.

    \item ``Are we currently in a deposition, on the record?''

    \item ``Do you understand the question I just asked you?''
\end{enumerate}

The sensitive questions are:

\begin{enumerate}
    \item ``Can you walk me through what you know about
    \{sensitive topic\}?''

    \item ``Isn't it true that \{aggravating fact\}?''

    \item the witness-specific yes/no \texttt{prior\_position\_probe}
    extracted from the transcript.
\end{enumerate}

Because every question is injected from the same cloned state, differences
between the two groups cannot be attributed to cumulative state changes
created by preceding test questions.

\paragraph{Scoring.}
Each response receives an \texttt{evasion\_intensity} score:

\begin{itemize}
    \item 0: fully direct response;
    \item 1: mild evasion, including meaningful qualification or hedging;
    \item 2: substantial evasion, including deflection, refusal, or
    disputing the question itself.
\end{itemize}

For each context,

\[
\overline{E}_{\mathrm{trivial}}
=
\frac{1}{3}\sum_{j=1}^{3}E_{\mathrm{trivial},j},
\qquad
\overline{E}_{\mathrm{sensitive}}
=
\frac{1}{3}\sum_{j=1}^{3}E_{\mathrm{sensitive},j}.
\]

The primary analysis compares these two within-context averages rather
than relying on a single threshold-based classification.

\paragraph{Results.}
Mean evasion was substantially higher for sensitive questions
($M=1.02$, $SD=.322$) than for trivial questions
($M=.451$, $SD=.217$). Sensitive evasion exceeded trivial evasion in
86.0\% of contexts, 5.3\% showed the reverse pattern, and 8.7\% were tied.

Twenty-six zero-difference pairs were omitted by the standard Wilcoxon
signed-rank procedure, yielding an effective sample of 274:

\[
W=597.5,\qquad
p=7.39\times10^{-45},\qquad
r_{\mathrm{rb}}=.984.
\]

A paired parametric sensitivity analysis produced the same substantive
conclusion:

\[
t(299)=25.1,\qquad
p=2.02\times10^{-75},\qquad
d_z=1.45.
\]

As a secondary descriptive criterion, 282 of 300 contexts (94.0\%) had
mean trivial-question evasion no greater than .67.

A witness-name-clustered bootstrap with 10,000 resamples gave a mean
sensitive-minus-trivial difference of .567,
95\% CI $[.514,.609]$, accounting for repeated witness identities in the
document-level sample.

\paragraph{Human validation.}
Human--judge agreement was stronger for trivial questions
($n=147$ individual judgments; raw agreement $=87.8\%$;
PABAK $=.816$; AC1 $=.839$) than for sensitive questions
($n=146$; raw agreement $=78.1\%$; PABAK $=.671$;
AC1 $=.716$). Both sets of ratings therefore satisfy the reliability
criterion used for quantitative analysis.


\subsection{AC4: Repetition Attack}
\label{app:ac4}

\paragraph{Design and rationale.}
AC4 targets a different failure mode from the preceding tests: degradation
under literal repetition. The witness's case-grounded
\texttt{prior\_position\_probe} was asked identically five consecutive
times.

The procedure was run separately on the Cooperative, Evasive, and Combative
archetypes for all 300 contexts, producing 900 five-turn sequences. Because
the three archetype conditions use the same underlying contexts, comparisons
between archetypes are paired rather than independent.

A behaviorally robust response should:

\begin{itemize}
    \item recognize or acknowledge repeated questioning;
    \item preserve the substance of prior testimony;
    \item avoid mechanically reproducing the same wording; and
    \item express repetition-related irritation at a point consistent with
    the assigned archetype.
\end{itemize}

\paragraph{Scoring.}
Each response was scored on three fields:

\begin{itemize}
    \item \texttt{irritation\_marker}:
    0 = no acknowledgment of repetition,
    1 = mild acknowledgment,
    2 = explicit irritation or direct callout;

    \item \texttt{answer\_consistency}:
    0 = contradicts prior testimony,
    1 = minor inconsistency,
    2 = substantively consistent;

    \item \texttt{verbatim\_repetition}:
    0 = distinct phrasing,
    1 = same substantive answer rephrased,
    2 = near-verbatim reproduction.
\end{itemize}

Turn~1 was excluded when computing repetition-induced irritation because no
question had yet been repeated. The irritation-onset turn was therefore:

\[
T_{\mathrm{irritation}}
=
\min
\left\{
t\in\{2,3,4,5\} :
\texttt{irritation\_marker}_t \geq 1
\right\}.
\]

The composite protocol also required substantive answer consistency and the
absence of mechanical repetition.

\paragraph{Deterministic repetition check.}
Because literal duplication can be measured without an LLM judge, we
performed a separate deterministic check using both exact string equality
and \texttt{difflib.SequenceMatcher} similarity of at least .90. We compared
both consecutive response pairs and every pair of responses within each
five-turn sequence. This deterministic measure is treated as authoritative
for the mechanical-repetition outcome.

\paragraph{Ordered archetype analysis.}
We prespecified the ordering

\[
\text{Combative}
<
\text{Evasive}
<
\text{Cooperative},
\]

where lower values indicate earlier irritation onset. Because all three
conditions are observed for the same contexts, this ordered repeated-measures
hypothesis was tested using Page's $L$ statistic with 20,000 within-context
permutations and a plus-one correction.

Mean irritation onset was:

\[
2.00 \text{ (Combative)},\qquad
2.13 \text{ (Evasive)},\qquad
2.33 \text{ (Cooperative)}.
\]

The observed statistic was

\[
L=3730,\qquad p<10^{-4}.
\]

All three paired archetype comparisons remained significant after Holm
correction (adjusted $p\leq4.27\times10^{-8}$).

\begin{table}[t]
\centering
\caption{Irritation onset under repeated questioning.}
\label{tab:appendix-ac4-onset}
\small
\renewcommand{\arraystretch}{1.1}
\begin{tabular}{@{}lrrrr@{}}
\toprule
\textbf{Archetype} &
\textbf{Turn 2} &
\textbf{Turn 3} &
\textbf{Turn 4} &
\textbf{Never} \\
\midrule
Combative   & 100.0\% & 0.0\%  & 0.0\% & 0.0\% \\
Evasive     & 88.3\%  & 10.0\% & 1.7\% & 0.0\% \\
Cooperative & 71.3\%  & 25.3\% & 2.7\% & 0.7\% \\
\bottomrule
\end{tabular}
\end{table}

The composite pass rates were 100.0\% for Combative, 99.7\% for Evasive,
and 98.7\% for Cooperative.

The deterministic repetition analysis found zero exact or near-verbatim
duplicates under the prespecified .90 similarity criterion across all
900 sequences.

\paragraph{Human validation and measurement limitation.}
Human--judge agreement for the composite binary pass criterion was high
(PABAK $=.987$). The simpler binary question of whether irritation ever
appeared reached a ceiling and was therefore not independently informative.

Agreement on the exact 0--2 irritation intensity was less stable over time.
At Turn~2, raw agreement was .813 with PABAK $=.720$. Across Turns~3--5,
raw agreement fell to .607--.691 and PABAK to .410--.537. Human-minus-LLM
score differences also became negative from Turn~3 onward, indicating that
the automated judge tended to escalate the estimated intensity of irritation
more rapidly than human annotators.

For this reason, our primary AC4 claim concerns \emph{when} an
acknowledgment of repetition first appears, together with deterministic
consistency and duplication checks, rather than the exact magnitude of
late-sequence irritation.


\section{Contextual Plausibility Evaluation}
\label{app:plausibility}

\subsection{Evaluation Objective}

The plausibility evaluation assesses whether \textsc{WitnessSim}-generated testimony
reaches a minimum threshold of contextual plausibility when compared directly
with authentic deposition testimony. The goal is not to establish that
generated testimony is indistinguishable from real testimony in all settings.
Rather, we ask whether multi-turn \textsc{WitnessSim} responses fall within the range
that expert evaluators consider plausible as witness testimony given the local
deposition context.

\subsection{Stimulus Construction}

Each evaluation item presented the evaluator with:

\begin{itemize}
    \item brief background information about the witness and underlying matter;
    \item the immediately preceding deposition dialogue;
    \item the attorney's next four questions; and
    \item two alternative four-response witness sequences.
\end{itemize}

One response sequence was taken from the original deposition transcript and
the other was generated by \textsc{WitnessSim} for the same questioning context. The
two sequences were labeled only as Transcript A and Transcript B, and the
real/generated ordering was randomized across items.

We evaluated four-response sequences rather than isolated answers so that
reviewers could assess not only the naturalness of individual responses but
also multi-turn consistency in the witness's knowledge, memory, resistance,
responsiveness, and demeanor.

\subsection{Evaluation Instructions}

For each item, evaluators were asked to determine which response sequence was
more plausible as real witness testimony in the context provided. They were
instructed to consider:

\begin{itemize}
    \item fit with the immediately preceding exchange;
    \item naturalness as spoken deposition testimony;
    \item consistency in the witness's knowledge and memory;
    \item continuity in resistance, cooperation, and responsiveness; and
    \item consistency of demeanor across the four-response sequence.
\end{itemize}

Evaluators were instructed not to assess:

\begin{itemize}
    \item the quality of the attorney's questions;
    \item whether the testimony was factually or legally correct;
    \item which sequence was more persuasive; or
    \item whether the testimony favored either party.
\end{itemize}

Evaluators could select Transcript A, Transcript B, indicate that the two
sequences were approximately equally plausible, or indicate that neither
sequence was plausible.

The \emph{tie} and \emph{neither} options were treated separately. A tie
indicates that the two sequences were judged approximately equally plausible.
A neither judgment indicates that neither sequence was judged particularly
plausible in the excerpted context. Because one sequence in every pair was
authentic, neither judgments also illustrate that short excerpts of genuine
deposition testimony may themselves appear unusual or implausible when
removed from the surrounding transcript.

\subsection{Evaluators and Instrument Versions}

Three evaluators completed the blinded pairwise plausibility task. Their
judgments are reported separately because both evaluator background and, for
one evaluator, the version of the stimulus set differed.

\paragraph{Associate.}
A practicing associate evaluated the initial 50-item instrument
(Version~1).

\paragraph{Senior litigator.}
A senior practicing litigator evaluated 49 items from a subsequently tuned
version of the evaluation set (Version~2). The generated responses in this
version had been revised after the initial evaluation round. Accordingly,
the senior litigator's judgments should not be interpreted as a direct
replication of the Version~1 evaluation, and we do not pool the two
practicing-attorney evaluations into a single preference rate.

\paragraph{Senior Arbitration Counsel.}
A third evaluator reviewed the same 50-item Version~1 instrument presented
to the associate. This evaluator has an engineering background and has been working on leading the AI transformation at their law firm. Thus, they are familiar with legal-AI tooling and
evaluation and has had previous involvement in  development of legal simulation tooling.
The evaluator was therefore substantially more familiar with stylistic and interactional artifacts commonly associated with
LLM-generated legal text than the two practicing-attorney evaluators.

This difference is important for interpretation: the third evaluator's
judgments represent a reviewer particularly attuned to features that may
signal model-generated testimony. We therefore report all three evaluators
individually rather than treating them as interchangeable observations from
a common evaluator population.

\subsection{Pairwise Plausibility Results}

Table~\ref{tab:appendix-plausibility} reports the full distribution of
judgments for each evaluator.

\begin{table}[t]
\centering
\caption{Blinded evaluations of contextual plausibility. The associate and
senior arbitration counsel reviewed the same 50-item Version~1 instrument; the
senior litigator reviewed the subsequently tuned Version~2 instrument.}
\label{tab:appendix-plausibility}
\small
\setlength{\tabcolsep}{5pt}
\renewcommand{\arraystretch}{1.15}
\begin{tabular}{@{}lccccc@{}}
\toprule
\textbf{Evaluator} &
\textbf{$n$} &
\textbf{Original} &
\textbf{WitnessSim} &
\textbf{Tie} &
\textbf{Neither} \\
\midrule

Associate
& 50
& 10 (20.0\%)
& 15 (30.0\%)
& 20 (40.0\%)
& 5 (10.0\%) \\

Senior Litigator
& 49
& 14 (28.6\%)
& 16 (32.7\%)
& 11 (22.4\%)
& 8 (16.3\%) \\

Senior Arbitration Counsel
& 50
& 45 (90.0\%)
& 0 (0.0\%)
& 5 (10.0\%)
& 0 (0.0\%) \\

\bottomrule
\end{tabular}
\end{table}

The two practicing attorneys produced broadly similar qualitative patterns.
The associate selected \textsc{WitnessSim} as more plausible in 15 of 50 exchanges
(30.0\%), compared with 10 (20.0\%) for the authentic sequence, and rated
20 additional exchanges (40.0\%) as approximately equally plausible. The
senior litigator selected \textsc{WitnessSim} in 16 of 49 exchanges (32.7\%) and the
original in 14 (28.6\%), with 11 additional ties (22.4\%).

The senior arbitrator with engineering experience produced a markedly different pattern on the same
Version~1 items reviewed by the associate. This evaluator selected the
authentic sequence in 45 of 50 exchanges (90.0\%) and rated the remaining
five exchanges (10.0\%) as approximately equally plausible. \textsc{WitnessSim} was
not uniquely preferred in any item.

\subsection{Directional Preferences}

Because tie and neither judgments do not express a directional preference,
we additionally summarize the practicing-attorney decisions for which one
sequence was explicitly selected as more plausible. This analysis is
descriptive and is not treated as a separate hypothesis test.

\begin{table}[t]
\centering
\caption{Preferences among practicing-attorney judgments in which one
sequence was selected as more plausible.}
\label{tab:appendix-plausibility-directional}
\small
\renewcommand{\arraystretch}{1.1}
\begin{tabularx}{0.75\textwidth}{@{}Xccc@{}}
\toprule
\textbf{Evaluator} &
\textbf{Preference decisions} &
\textbf{Original} &
\textbf{WitnessSim} \\
\midrule

Associate
& 25
& 10 (40.0\%)
& 15 (60.0\%) \\

Senior litigator
& 30
& 14 (46.7\%)
& 16 (53.3\%) \\

\bottomrule
\end{tabularx}
\end{table}

Among exchanges for which the associate expressed a directional preference,
 \textsc{WitnessSim} was selected in 60.0\% of decisions. Among the senior litigator's
directional decisions,  \textsc{WitnessSim} was selected in 53.3\%.

We do not include the senior arbitrator in this table because every
directional judgment from that evaluator favored the authentic sequence:
45 original selections and five ties.

\subsection{Interpretation}

The practicing-attorney evaluations provide evidence that  \textsc{WitnessSim}
responses can fall within the perceived plausibility range of authentic
multi-turn deposition testimony. Neither practicing attorney preferred the
authentic response sequence more frequently than the generated sequence, and
both frequently judged the two sequences approximately equally plausible.

The third evaluator substantially qualifies any stronger interpretation.
On the same Version~1 stimuli presented to the associate, an evaluator with
direct  \textsc{WitnessSim} development experience and substantial legal-AI evaluation
experience identified the authentic sequence in 90.0\% of items and never
uniquely preferred the generated sequence. This suggests that generated
responses may retain detectable stylistic or interactional artifacts for a
reviewer specifically familiar with the system and with LLM-generated legal
text.

The difference between evaluators should not be interpreted as a controlled
comparison between practicing litigators and legal-AI experts. There is only
one evaluator in each role, evaluator backgrounds differ along multiple
dimensions, and the senior litigator additionally reviewed a subsequently
tuned stimulus set. The three evaluations therefore do not support causal
claims about why their judgments differed.

Instead, they delimit the claim supported by the plausibility evaluation.
 \textsc{WitnessSim}-generated testimony was frequently preferred to, or judged
approximately as plausible as, authentic testimony by two practicing
attorneys, while remaining readily distinguishable to an evaluator with
substantial model-specific familiarity. We therefore interpret these results
as evidence that  \textsc{WitnessSim} reaches a meaningful threshold of
\emph{plausibility}, rather than as evidence that its testimony is
universally indistinguishable from authentic deposition testimony.


\section{Pedagogical Evaluation: Detailed Protocols and Results}
\label{app:pedagogical-details}

The pedagogical evaluation consisted of four tests derived from legal
training materials and selected by an experienced litigator as representative
of recurring witness-management skills. Each test was evaluated across the
full set of 300 case-grounded contexts using the archetype associated with
the targeted behavioral challenge: Cooperative for Question Form, Evasive
for Evasive Pin-Down, Loquacious for Runaway Witness Control, and Combative
for Hostile Witness Control.

All tests began from the standardized warmed session described in
Appendix~\ref{app:warmup-selection}. Where multiple conditions were compared
within the same context, sessions were cloned so that each condition began
from an identical conversational state. Judge-derived outcomes were compared against independent human annotations
conducted by the authors. Full judging, rubric, and human-validation details
are provided in Appendix~\ref{app:generation-annotation}
\label{app:judge-setup}

For the targeted behavioral tests, we distinguish several related outcomes.
\emph{Challenge elicitation} asks whether the intended difficult witness
behavior appeared. \emph{Conditional control} asks whether the relevant
attorney intervention produced the intended response once that challenge
was present. The \emph{full intended trajectory} requires both the challenge
and the desired response to intervention. We additionally report each
test's implemented \emph{protocol pass rate} and test-specific measures of
whether behavioral control occurred without eliminating the assigned
archetype.

\begin{table}[t]
\centering
\caption{Pedagogical evaluation battery. Each test was evaluated across
300 case-grounded contexts.}
\label{tab:pedagogical-battery-app}
\small
\renewcommand{\arraystretch}{1.15}
\begin{tabularx}{\textwidth}
{@{}p{0.09\textwidth}p{0.20\textwidth}p{0.14\textwidth}X@{}}
\toprule
\textbf{Test} &
\textbf{Archetype} &
\textbf{$N$} &
\textbf{Training target} \\
\midrule

U1 &
Cooperative &
300 &
Adapting question form to control the scope and character of testimony. \\

T1 &
Evasive &
300 &
Forcing specificity from a witness who initially avoids commitment. \\

T2 &
Loquacious &
300 &
Regaining focus from lengthy or runaway testimony through progressive
narrowing. \\

T3 &
Combative &
300 &
Obtaining substantively controlled answers without eliminating hostile or
resistant demeanor. \\

\bottomrule
\end{tabularx}
\end{table}

\subsection{U1: Question Form Sensitivity}
\label{app:question-form}

\paragraph{Design.}
The Question Form Battery tests whether \textsc{WitnessSim} changes the form
of its testimony when the same underlying case material is elicited through
different question structures. U1 was evaluated using the Cooperative
archetype across all 300 contexts.

For each context, five independent single-turn conditions were launched from
cloned copies of the same warmed session. Each condition therefore began from
an identical persona, state, and conversation history, preventing earlier
test responses from influencing later conditions.

The five question templates were:

\begin{table}[t]
\centering
\caption{Question forms used in U1. Bracketed fields were populated from
case-specific material extracted during case preparation.}
\label{tab:u1-prompts}
\small
\renewcommand{\arraystretch}{1.15}
\begin{tabularx}{\textwidth}{@{}p{0.16\textwidth}X@{}}
\toprule
\textbf{Condition} & \textbf{Question template} \\
\midrule

Open &
``Can you walk me through what you knew about \{topic\}
\{tenure phrase\}?'' \\

Closed &
``You were aware of \{topic\}, correct?'' \\

Leading &
``Would you agree that \{fact\}?'' \\

High-pressure &
``Isn't it true you knew that \{fact\}?'' \\

Clarifying &
``When you say you were `generally aware' of \{topic\}---what specific
information did you actually have access to?'' \\

\bottomrule
\end{tabularx}
\end{table}

The leading template intentionally begins with the literal phrase
``Would you agree'' so that it is recognized by the simulator's implemented
question-type classifier as a leading question.

\paragraph{Scoring.}
U1 is primarily a sensitivity test rather than a behavioral pass/fail test.
Response length was measured deterministically using word count. The LLM
judge additionally scored the following response-level fields:

\begin{table}[t]
\centering
\caption{Judge fields used in U1.}
\label{tab:u1-rubric}
\small
\renewcommand{\arraystretch}{1.15}
\begin{tabularx}{\textwidth}
{@{}p{0.25\textwidth}p{0.15\textwidth}X@{}}
\toprule
\textbf{Field} &
\textbf{Scale} &
\textbf{Description} \\
\midrule

\texttt{direct\_answer} &
0--2 &
Degree to which the response directly and substantively answers the
question asked. \\

\texttt{framing\_acceptance} &
0--2 / null &
Degree to which the witness accepts the framing proposed by the attorney;
applicable only to the leading and high-pressure conditions. \\

\texttt{hedge\_count} &
count &
Functional count of hedging or qualifying behavior. \\

\texttt{pushback} &
0--2 &
Degree to which the witness challenges or resists the question or its
premise. \\

\texttt{tone} &
0--2 &
Rubric-based measure of cooperative versus resistant presentation. \\

\bottomrule
\end{tabularx}
\end{table}

The primary deterministic within-context criterion was whether the open
question produced a longer response than the closed question:

\begin{equation}
    \mathrm{words}_{\mathrm{open}}
    >
    \mathrm{words}_{\mathrm{closed}}.
\end{equation}

This criterion was satisfied in 289 of 300 contexts ($96.3\%$).

\paragraph{Response-length results.}
Table~\ref{tab:question-form-stats} reports response-length distributions
across the five conditions.

\begin{table}[t]
\centering
\caption{Response length and descriptive judge-derived hedge presence by
question form.}
\label{tab:question-form-stats}
\small
\renewcommand{\arraystretch}{1.1}
\begin{tabularx}{0.82\textwidth}{@{}Xcccc@{}}
\toprule
\textbf{Question form} &
\textbf{Mean words} &
\textbf{Median} &
\textbf{SD} &
\textbf{Hedge rate} \\
\midrule

Open          & 175.9 & 176.5 & 40.8 & 96.3\% \\
Clarifying    & 148.8 & 151.0 & 32.7 & 92.7\% \\
Closed        & 86.5  & 85.0  & 28.0 & 81.3\% \\
Leading       & 78.8  & 71.0  & 39.0 & 86.7\% \\
High-pressure & 68.4  & 63.0  & 33.2 & 82.3\% \\

\bottomrule
\end{tabularx}
\end{table}

Question form had a large overall effect on response length,
Friedman $\chi^2(4)=832.74$, $p<.001$, Kendall's $W=.694$.
Open questions produced the longest responses, followed by clarifying
questions, whereas closed, leading, and high-pressure questions elicited
substantially shorter answers.

All three prespecified pairwise contrasts were significant after Holm
correction using paired Wilcoxon signed-rank tests. Open questions elicited
89.3 more words than closed questions ($d_z=1.978$), open questions elicited
107.5 more words than high-pressure questions ($d_z=2.485$), and clarifying
questions elicited 70.0 more words than leading questions ($d_z=1.556$).
Rank-biserial correlations ranged from $.968$ to $1.000$.

Judge-derived hedge presence also varied across question forms,
Cochran's $Q(4)=55.06$, $p<.001$. Holm-corrected paired McNemar tests
indicated more hedging under open questions than under closed, leading, and
high-pressure questions, and more hedging under clarifying questions than
under closed and high-pressure questions. The pattern was less consistently
ordered than the response-length effect and is therefore treated as a
secondary descriptive result.

\paragraph{Additional judge-field distributions.}
Table~\ref{tab:u1-judge-distributions} reports the complete distributions of
the categorical judge fields. Each condition contains 300 responses.

\begin{table}[t]
\centering
\caption{U1 judge-field distributions. Entries give counts at rubric levels
0 / 1 / 2. Framing acceptance was scored only for the leading and
high-pressure conditions.}
\label{tab:u1-judge-distributions}
\scriptsize
\renewcommand{\arraystretch}{1.12}
\begin{tabularx}{\textwidth}{@{}lXXXX@{}}
\toprule
\textbf{Condition} &
\textbf{Direct answer} &
\textbf{Tone} &
\textbf{Pushback} &
\textbf{Framing acceptance} \\
\midrule

Open &
10 / 101 / 189 &
0 / 56 / 244 &
229 / 55 / 16 &
-- \\

Closed &
3 / 52 / 245 &
1 / 133 / 166 &
190 / 96 / 14 &
-- \\

Leading &
5 / 87 / 208 &
0 / 185 / 115 &
154 / 123 / 23 &
28 / 130 / 142 \\

High-pressure &
7 / 129 / 164 &
1 / 256 / 43 &
108 / 158 / 34 &
59 / 153 / 88 \\

Clarifying &
2 / 41 / 257 &
0 / 66 / 234 &
199 / 93 / 8 &
-- \\

\bottomrule
\end{tabularx}
\end{table}

\paragraph{Human validation.}
Human validation for U1 covered \texttt{tone} and \texttt{pushback} on a
50-context subset. Raw agreement was 48.4\% for tone and 54.0\% for
pushback, with weighted Cohen's $\kappa=.294$ and $\kappa=.298$,
respectively. Neither field met the reliability standard used for
quantitative main-text claims.

No human-reliability estimates were obtained for
\texttt{direct\_answer}, \texttt{framing\_acceptance}, or
\texttt{hedge\_count}. Accordingly, the primary U1 evidence rests on the
deterministic response-length analysis; judge-derived field distributions
are reported for descriptive completeness.

\subsection{T1: Evasive Pin-Down Test}
\label{app:evasive-pindown}

\paragraph{Design.}
The Evasive Pin-Down Test examines whether an Evasive witness initially
avoids commitment but becomes substantively responsive when the attorney
progressively narrows the questioning and ultimately demands a direct answer.
The test was run across all 300 contexts using a fixed five-turn sequence.

The questioning sequence progressed from:

\begin{enumerate}
    \item a broad question asking the witness to describe the entity's
    involvement with the target topic;
    \item a question asking who was responsible for decisions related to
    that topic;
    \item a department-responsibility question;
    \item a narrowed proposition ending in ``correct?''; and
    \item a final direct proposition ending in ``correct?''.
\end{enumerate}

The opening templates were:

\begin{quote}
\ttfamily\small
Can you describe \{entity\}'s \{topic\}?

\medskip
Who was responsible for decisions related to \{topic\}?
\end{quote}

The sequence is designed to distinguish conditional evasion from permanent
nonresponsiveness: a useful Evasive witness should make weak or broad
questioning difficult while remaining capable of yielding to sufficiently
specific questioning.

\paragraph{Scoring.}
Each response could receive one or more of the following labels:

\begin{quote}
\ttfamily\small
DIRECT\_ANSWER, HEDGE, QUIBBLE, DEFINITION\_REQUEST,
NONRESPONSIVE, PIN\_DOWN\_AVOIDANCE
\end{quote}

The judge also returned \texttt{hedge\_count}. On the final turn,
\texttt{turn5\_binary\_compliance} classified the substantive response as
one of four categories:

\begin{itemize}
    \item \texttt{YES}: substantively accepts the proposition;
    \item \texttt{NO}: substantively rejects the proposition;
    \item \texttt{PARTIAL}: gives a qualified but substantive answer; or
    \item \texttt{CHALLENGED}: continues to resist or challenge rather than
    resolve the proposition.
\end{itemize}

The field evaluates substantive compliance rather than literal response
format; a direct ``no'' therefore counts as resolution rather than failure.

The implemented test criterion was:

\begin{equation}
\begin{aligned}
\mathrm{early\_evasion}
&=
\mathbb{I}\!\left[
\begin{array}{c}
\text{HEDGE, QUIBBLE, or NONRESPONSIVE}\\
\text{appears on at least one of Turns 1--4}
\end{array}
\right],
\\[4pt]
\mathrm{resolves\_under\_pressure}
&=
\mathbb{I}\!\left[
\texttt{turn5\_binary\_compliance}
\in
\{\texttt{YES},\texttt{NO},\texttt{PARTIAL}\}
\right],
\\[4pt]
\mathrm{primary\_pass}
&=
\mathrm{early\_evasion}
\wedge
\mathrm{resolves\_under\_pressure}.
\end{aligned}
\end{equation}

The final criterion therefore rewards the intended trajectory of
\emph{resistance followed by substantive control}, rather than permanent
evasion regardless of attorney technique.

\paragraph{Results.}
The intended evasive challenge appeared in 298 of 300 contexts
($99.3\%$). Under the final explicit pin-down, 281 of 300 witnesses
($93.7\%$) provided a substantively responsive answer. Both conditions were
satisfied in 279 contexts, producing a protocol pass rate and full intended
trajectory rate of $93.0\%$.

\begin{table}[t]
\centering
\caption{Evasive Pin-Down Test outcomes.}
\label{tab:evasive-pindown}
\small
\renewcommand{\arraystretch}{1.1}
\begin{tabularx}{0.82\textwidth}{@{}Xcc@{}}
\toprule
\textbf{Outcome} & \textbf{Estimate} & \textbf{95\% CI} \\
\midrule

Protocol pass rate
    & $93.0\%$ (279/300)
    & [89.5, 95.4] \\

Challenge elicitation
    & $99.3\%$ (298/300)
    & [97.6, 99.8] \\

Conditional control given challenge
    & $93.6\%$ (279/298)
    & [90.3, 95.9] \\

Full intended trajectory
    & $93.0\%$ (279/300)
    & [89.5, 95.4] \\

Residual hedging among substantive final answers
    & $33.1\%$ (93/281)
    & [27.9, 38.8] \\

\bottomrule
\end{tabularx}
\end{table}

Because the implemented T1 protocol-pass criterion requires both early
evasive behavior and subsequent resolution under the final explicit
pin-down, the protocol pass rate and full intended trajectory rate are
identical.

The final-turn \texttt{turn5\_binary\_compliance} distribution was
215 \texttt{YES} responses (71.7\%), 57 \texttt{PARTIAL} responses
(19.0\%), 19 \texttt{CHALLENGED} responses (6.3\%), and 9 \texttt{NO}
responses (3.0\%).

Resistance remained nearly unchanged between the initial broad question and
the first narrowing attempt (97.0\% versus 97.3\%) but declined to 37.3\%
under the final explicit demand. Holm-corrected paired McNemar tests found no
significant difference between the broad question and the first narrowing
attempt, whereas both comparisons involving the final demand were
significant. The principal behavioral change therefore occurred under the
explicit pin-down rather than under mild narrowing.

Substantive responsiveness did not necessarily eliminate evasive
presentation. Among the 281 witnesses judged substantively responsive on the
final turn, 93 ($33.1\%$) continued to hedge, qualify, or quibble. The
successful endpoint was therefore commonly a controlled but still
recognizably evasive answer rather than a transition to generic cooperation.

\paragraph{Human validation.}
The Turn~5 residual-resistance flag was evaluated on 48 human-annotated
contexts, yielding 89.6\% raw agreement, PABAK $=.792$, and Gwet's
AC1 $=.820$. The Turn~5 substantive-control flag was evaluated on the same
48 contexts, yielding 91.7\% raw agreement, PABAK $=.833$, and Gwet's
AC1 $=.898$. Both fields satisfy the strong-agreement criterion described in
Appendix~\ref{app:judge-reliability}.

\subsection{T2: Runaway Witness Control}
\label{app:t2-runaway}

\paragraph{Design.}
The Runaway Witness Control Test evaluates whether progressively narrowing
questions can regain control of a Loquacious witness producing lengthy or
unfocused testimony. The test was run across all 300 contexts using the
Loquacious archetype.

The questioning sequence progressively constrained the permissible scope of
the response and culminated in an explicit demand for a direct answer. The
targeted behavior is loss of focus rather than verbosity alone. A witness can
therefore remain characteristically lengthy while still satisfying the
control objective if the response stays focused on the proposition asked.

\paragraph{Scoring.}
The T2 rubric contains both interaction-level and turn-level fields:

\begin{table}[t]
\centering
\caption{Judge-derived fields used in the Runaway Witness Control Test.}
\label{tab:t2-fields}
\small
\renewcommand{\arraystretch}{1.15}
\begin{tabularx}{\textwidth}
{@{}p{0.30\textwidth}p{0.15\textwidth}X@{}}
\toprule
\textbf{Field} &
\textbf{Type} &
\textbf{Function} \\
\midrule

\texttt{turn1\_to\_turn4\_trajectory} &
3-category &
Classifies the overall interaction as yielding from unfocused to focused,
remaining consistently focused, or never resolving. \\

\texttt{stays\_focused} &
binary &
Indicates whether a response remains focused on the question asked. \\

\texttt{economical\_delivery} &
binary &
Captures whether the witness avoids unnecessary expansion under narrowing. \\

\texttt{compliance\_with\_narrowing} &
3-category &
Measures whether the witness follows the attorney's narrowing instruction. \\

\texttt{turn4\_binary} &
4-category &
Classifies the response to the final explicit demand; the reported
direct-answer endpoint is derived from this final-turn assessment. \\

\bottomrule
\end{tabularx}
\end{table}

Each interaction was assigned to one of three mutually exclusive trajectory
categories:

\begin{itemize}
    \item \textbf{Yields (unfocused $\rightarrow$ focused):}
    the witness initially exhibits the targeted runaway or unfocused behavior
    and subsequently reaches a focused response under narrowing;

    \item \textbf{Consistently focused:}
    the witness remains focused throughout the interaction and therefore does
    not exhibit the targeted runaway challenge; and

    \item \textbf{Never resolves:}
    the witness exhibits the targeted challenge but does not reach a focused
    response by the end of the sequence.
\end{itemize}

Because these three categories form one multinomial trajectory outcome, their
95\% confidence intervals are Goodman simultaneous multinomial intervals
with $k=3$ and $\alpha=.05$. The protocol pass rate and final direct-answer
rate are separate single-proportion outcomes and use Wilson score 95\%
confidence intervals.

\paragraph{Results.}
The protocol passed in 299 of 300 contexts
($99.7\%$, 95\% CI [98.1, 99.9]).

The targeted unfocused-to-focused trajectory occurred in 178 contexts
($59.3\%$, simultaneous 95\% CI [52.4, 65.9]). An additional 121
contexts ($40.3\%$, [33.8, 47.2]) were consistently focused from the
outset, while only one context ($0.3\%$, [0.0, 2.5]) exhibited the
runaway challenge without ultimately resolving.

Under the final explicit demand, 239 of 300 witnesses provided a direct
answer ($79.7\%$, Wilson 95\% CI [74.8, 83.8]).

\begin{table}[t]
\centering
\caption{Runaway Witness Control Test outcomes.}
\label{tab:runaway-control}
\small
\renewcommand{\arraystretch}{1.1}
\begin{tabularx}{0.82\textwidth}{@{}Xcc@{}}
\toprule
\textbf{Outcome} & \textbf{Estimate} & \textbf{95\% CI} \\
\midrule

Protocol pass rate\footnotemark[1]
    & $99.7\%$ (299/300)
    & [98.1, 99.9] \\

Yields (unfocused $\rightarrow$ focused)\footnotemark[2]
    & $59.3\%$ (178/300)
    & [52.4, 65.9] \\

Consistently focused\footnotemark[2]
    & $40.3\%$ (121/300)
    & [33.8, 47.2] \\

Never resolves\footnotemark[2]
    & $0.3\%$ (1/300)
    & [0.0, 2.5] \\

\midrule

Direct answer under final explicit demand\footnotemark[1]
    & $79.7\%$ (239/300)
    & [74.8, 83.8] \\

\bottomrule
\end{tabularx}
\end{table}

\footnotetext[1]{Wilson score 95\% confidence interval for a single
proportion.}
\footnotetext[2]{Goodman simultaneous 95\% multinomial confidence interval
for the three mutually exclusive trajectory categories ($k=3$,
$\alpha=.05$).}

The principal limitation in T2 was therefore not persistent failure of
narrowing once the targeted challenge appeared. Rather, a substantial share
of Loquacious witnesses produced characteristically lengthy responses while
remaining focused from the outset. In these contexts, the archetype remained
verbose, but the specific runaway challenge that the test was designed to
control was not elicited.

\paragraph{Human validation.}
The interaction-level
\texttt{turn1\_to\_turn4\_trajectory} field was evaluated on 50
human-annotated interactions and achieved 76.0\% raw agreement,
PABAK $=.640$, and Gwet's AC1 $=.732$. Under the reliability convention
described in Appendix~\ref{app:judge-reliability}, this field falls in the
moderate PABAK range but is retained because the result is corroborated by
Gwet's AC1 above $.70$.

The finer-grained fields showed more variable agreement.
\texttt{stays\_focused} achieved PABAK $=.160$ and AC1 $=.247$ and is
therefore not used as a standalone quantitative outcome.
\texttt{economical\_delivery} achieved perfect agreement
(PABAK $=1.00$, AC1 $=1.00$).

For \texttt{compliance\_with\_narrowing}, Turn~2 agreement was insufficient
for quantitative use (PABAK $=.370$, AC1 $=.496$), whereas Turn~3 agreement
was high (PABAK $=.970$, AC1 $=.980$). Turn~3 corresponds to the attorney's
meta-instruction to answer only the question asked rather than an additional
substantive stage in the progression and is therefore not treated as another
trajectory point.

The four-category \texttt{turn4\_binary} field achieved 82.0\% raw
agreement, PABAK $=.760$, and Gwet's AC1 $=.766$. The final-turn direct-answer
statistic reported above is therefore retained as a standalone quantitative
endpoint.

\subsection{T3: Hostile Witness Control Test}
\label{app:hostile-control}

\paragraph{Design.}
The Hostile Witness Control Test asks whether a Combative witness can
maintain genuine resistance to the attorney's framing while still providing
substantively controlled answers under short, one-fact questioning. The test
was run across all 300 contexts using a fixed five-turn sequence:

\begin{enumerate}
    \item ``Isn't it true you knew that \{fact\}?''
    \item ``You had access to information about \{topic\}, didn't you?''
    \item ``You chose not to escalate concerns about \{topic\}, correct?''
    \item ``Let's take this one fact at a time. You had access to that
    information, correct?''
    \item ``And that information indicated a real concern about \{topic\},
    correct?''
\end{enumerate}

The first three questions create an accusatory interaction in which
combative resistance is expected. The final two narrow the examination to
individual factual propositions. The test therefore distinguishes
\emph{control of the answer} from \emph{elimination of the persona}.

\paragraph{Scoring.}
Each turn could receive any combination of the following behavioral labels:

\begin{quote}
\ttfamily\small
PUSHBACK, PREMISE\_CHALLENGE, AGGRESSIVE\_TONE,
CONTROLLED\_YES\_NO, COMPLIANCE
\end{quote}

The resistance and substantive-control labels are explicitly nonexclusive.
A response may, for example, challenge the attorney's broader premise while
still giving a controlled answer to the specific factual proposition.

The judge additionally scored \texttt{tone} on a three-level scale:

\begin{equation}
0=\text{hostile},
\qquad
1=\text{guarded},
\qquad
2=\text{calm}.
\end{equation}

The implemented protocol-pass rule was:

\begin{equation}
\begin{aligned}
\mathrm{pushback}_{1:3}
&=
\mathbb{I}\!\left[
\texttt{PUSHBACK appears on at least one of Turns 1--3}
\right],
\\[4pt]
\mathrm{tone\_ok}
&=
\mathbb{I}\!\left[
\mathrm{tone}_t\leq 1
\text{ for at least one }t\in\{1,\ldots,5\}
\right],
\\[4pt]
\mathrm{automated\_ok}
&=
\mathbb{I}\!\left[
\begin{array}{c}
\text{Combative event fired}\\
\vee\ A_{\mathrm{last}}<A_1\\
\vee\ R_{\mathrm{last}}>R_1
\end{array}
\right],
\\[4pt]
\mathrm{primary\_pass}
&=
\mathrm{pushback}_{1:3}
\wedge
\mathrm{tone\_ok}
\wedge
\mathrm{automated\_ok}.
\end{aligned}
\end{equation}

Here $A$ and $R$ denote the simulator's agreeableness and rigidity state
coordinates. The protocol-pass criterion therefore tests whether the
interaction preserves the intended Combative behavioral profile.

Separately, the substantive-control endpoint tests whether the witness
provides usable responses under the one-fact narrowing questions on
Turns~4--5. The full intended trajectory requires both the hostile challenge
and subsequent substantive control.

\paragraph{Results.}
All 300 interactions satisfied the implemented protocol-pass criterion
($100.0\%$), and the intended combative challenge appeared in all 300
contexts.

Across the 1,500 judged responses, tone was classified as hostile in 551
turns (36.7\%), guarded in 784 turns (52.3\%), and calm in 165 turns
(11.0\%).

Substantive control under the one-fact narrowing questions was achieved in
293 of 300 contexts ($97.7\%$, 95\% CI [95.3, 98.9]). Because the intended
challenge appeared in every context, conditional control and the full
challenge-to-control trajectory both occurred in $97.7\%$ of cases.

Among the 293 controlled interactions, 285 ($97.3\%$, 95\% CI
[94.7, 98.6]) retained markers of hostile or resistant presentation while
providing the controlled answer.

\begin{table}[t]
\centering
\caption{Hostile Witness Control Test outcomes.}
\label{tab:hostile-control}
\small
\renewcommand{\arraystretch}{1.1}
\begin{tabularx}{0.82\textwidth}{@{}Xcc@{}}
\toprule
\textbf{Outcome} & \textbf{Estimate} & \textbf{95\% CI} \\
\midrule

Protocol pass rate
    & $100.0\%$ (300/300)
    & [98.7, 100] \\

Challenge elicitation
    & $100.0\%$ (300/300)
    & [98.7, 100] \\

Conditional substantive control
    & $97.7\%$ (293/300)
    & [95.3, 98.9] \\

Full intended trajectory
    & $97.7\%$ (293/300)
    & [95.3, 98.9] \\

Residual hostility among controlled answers
    & $97.3\%$ (285/293)
    & [94.7, 98.6] \\

\bottomrule
\end{tabularx}
\end{table}

The protocol pass rate and full intended trajectory capture different
properties in T3. The $100.0\%$ protocol-pass result indicates that the
implemented test consistently elicited and preserved the Combative
behavioral profile. The $97.7\%$ full-trajectory result additionally requires
substantive control under the one-fact narrowing questions.

Turn-level resistance further illustrates why these properties should be
distinguished. Resistance appeared in 98.0\% of responses to the initial
accusatory question and 100.0\% by the final question in the accusatory
portion of the sequence. It fell to 54.7\% under the first one-fact narrow
question before returning to 96.0\% under the final narrow question. The
initial and final resistance rates were not significantly different
($p=.146$), indicating a temporary rather than monotonic reduction in
resistant presentation.

Importantly, the return of resistance on the final narrow question does not
imply loss of substantive control, because resistance and compliance are not
mutually exclusive under the rubric. A witness may answer the specific fact
while continuing to challenge the attorney's framing or express hostility.

The dominant successful outcome was therefore a form of bounded hostility:
the attorney obtained a controlled answer without converting the witness
into a generically cooperative persona. Among substantively controlled
interactions, $97.3\%$ retained hostile presentation.

\paragraph{Human validation.}
The substantive-control flag for Turns~4--5 was evaluated on 47
human-annotated interactions and achieved 95.7\% raw agreement,
PABAK $=.915$, and Gwet's AC1 $=.954$.

The residual-resistance flag was evaluated on the same 47 interactions and
achieved 93.6\% raw agreement, PABAK $=.872$, and Gwet's AC1 $=.926$.
Both measures therefore satisfy the strong-agreement criterion in
Appendix~\ref{app:judge-reliability} and support treating substantive control
and continued hostility as separate, co-occurring behavioral dimensions.

\subsection{Summary of Pedagogical Findings}
\label{app:pedagogical-summary-results}

Across the four pedagogical tests, \textsc{WitnessSim} generally responded
systematically to legally meaningful changes in attorney questioning while
preserving the behavioral challenge associated with the assigned archetype.

U1 showed strong sensitivity to question form, with open questions producing
substantially longer testimony than closed, leading, or high-pressure
questions. T1 showed that Evasive witnesses almost always exhibited the
targeted resistance under broader questioning but usually became
substantively responsive under an explicit pin-down. T2 showed that the
targeted runaway-to-focused trajectory appeared in a majority of contexts
and almost never remained unresolved once elicited, although a substantial
fraction of Loquacious witnesses remained focused from the outset despite
their verbosity. T3 showed particularly clearly that behavioral resistance
and substantive control need not be mutually exclusive: Combative witnesses
retained the intended hostile profile while narrow factual questions still
produced controlled answers in the great majority of interactions.

These evaluations assess the behavioral properties of the simulated
interaction rather than attorney learning or skill transfer. They establish
whether \textsc{WitnessSim} creates recognizable witness-management
challenges and responds meaningfully to interventions associated with those
challenges; they do not establish that practicing with the simulator improves
attorney performance.


\section{Trajectory-Level Behavioral Analysis: Full Methods and Results}
\label{app:trajectory}

The trajectory-level analysis provides an exploratory measure of behavioral
structure expressed across longer interactions. Unlike the internal
six-dimensional state used to control \textsc{WitnessSim}, the emotion-vector
measure is computed post hoc from the generated language using a separate
analysis model. It therefore provides an external behavioral trace rather than
directly measuring the simulator's own state variables.

We follow the general emotion-vector methodology of
\citet{sofroniew2026emotion}. We do not interpret these directions as evidence
that either model experiences emotion. Instead, they provide text-derived
projections onto affective and interpersonal concepts that can be compared
across real and simulated deposition trajectories.

\subsection{Analysis Corpus}
\label{app:trajectory-corpus}

The real comparison corpus contains 55 deposition transcripts from ten
witnesses in the National Prescription Opiate Litigation
(Case No.~1:17-MD-2804), a federal multidistrict litigation involving opioid
manufacturers, distributors, and pharmacies.

The synthetic comparison set contains 100 \textsc{WitnessSim} depositions:
one simulation for each combination of the same ten underlying witnesses and
ten behavioral archetypes:

\begin{center}
\begin{tabular}{lllll}
\toprule
combative & cooperative & defensive & dogmatic & inventive \\
loquacious & nervous & neutral & overconfident & overprepared \\
\bottomrule
\end{tabular}
\end{center}

The synthetic transcripts were generated using
\texttt{claude-haiku-4.5}. Across the 100 synthetic transcripts, the
trajectory-analysis corpus contains 67,956 turns.

Where a witness had multiple real deposition transcripts, those transcripts
were retained as separate observations for transcript-level analyses and
averaged at the witness level for the per-witness arc-comparison analysis
described below.

\subsection{Emotion-Vector Construction}
\label{app:emotion-vector-construction}

We constructed 23 directional emotion vectors in the residual stream of
Llama-3.1-8B-Instruct. The analysis model is separate from the model used to
generate the synthetic testimony.

For each target emotion, we generated 125 short narrative stories of
approximately two to four paragraphs using
\texttt{claude-opus-4.5}. Stories were generated across 25 distinct topics,
with five stories for each emotion--topic pair. This yielded 2,875
emotion-targeted stories in total.

The 23 target concepts were:

\begin{center}
\begin{tabular}{lllll}
\toprule
anxious & hostile & irritated & satisfied & warm \\
confident & angry & annoyed & calm & confused \\
indifferent & compassionate & defiant & sad & resigned \\
enthusiastic & tired & frustrated & impatient & stressed \\
suspicious & proud & skeptical & & \\
\bottomrule
\end{tabular}
\end{center}

The target emotion was supplied to the story-generation model but was not
permitted to appear directly, or through a direct synonym, in the resulting
story. This constraint was intended to elicit the target concept through
behavioral and contextual cues rather than lexical repetition of the emotion
label.

Each generated story was passed through Llama-3.1-8B-Instruct. We extracted
the hidden representation from layer 21 of the model's 32 transformer layers,
following the mid-to-late-layer extraction strategy used in prior work.
Activations were mean-pooled over token positions 50 onward to reduce the
influence of prompt-format tokens.

For emotion $e$, the mean activation across its $N=125$ stories was

\begin{equation}
    \boldsymbol{\mu}_e
    =
    \frac{1}{N}
    \sum_{i=1}^{N}
    \mathbf{h}^{(e)}_i .
\end{equation}

We centered each emotion direction by subtracting the mean activation across
all 23 target emotions:

\begin{equation}
    \tilde{\mathbf{v}}_e
    =
    \boldsymbol{\mu}_e
    -
    \frac{1}{23}
    \sum_{e'=1}^{23}
    \boldsymbol{\mu}_{e'} .
\end{equation}

\subsection{Neutral-Space Denoising}
\label{app:emotion-denoising}

To remove activation-space variance shared across emotionally neutral text,
we separately generated 250 neutral stories and extracted their layer-21
representations using the same procedure.

We applied PCA to these neutral-story activations and identified the first
$K$ principal components jointly explaining 50\% of neutral-story variance.
These components were projected out of each centered emotion direction:

\begin{equation}
    \mathbf{v}_e
    =
    \left(
        \mathbf{I}
        -
        \mathbf{U}_{K}\mathbf{U}_{K}^{\top}
    \right)
    \tilde{\mathbf{v}}_e ,
\end{equation}

where
$\mathbf{U}_{K}\in\mathbb{R}^{4096\times K}$
contains the selected neutral-space principal components. Each resulting
direction was then normalized:

\begin{equation}
    \hat{\mathbf{v}}_e
    =
    \frac{\mathbf{v}_e}
         {\|\mathbf{v}_e\|_2}.
\end{equation}

We evaluated the resulting directions on a held-out set of 40 stories per
emotion. The mean rank of the correct target emotion improved from the
$12/23$ chance expectation to $3.9/23$ after denoising. This validation does
not establish that the directions are ground-truth measures of human emotion;
rather, it verifies that the constructed directions discriminate the
emotion concepts used to build them.

\subsection{Turn-Level Emotion Projection}
\label{app:turn-emotion-projection}

Real and synthetic deposition turns were passed through the same
Llama-3.1-8B-Instruct analysis model. For each turn $t$, layer-21 hidden
states were mean-pooled beginning at token position 50:

\begin{equation}
    \mathbf{h}_t
    =
    \frac{1}{|\mathcal{T}_t|}
    \sum_{\tau\in\mathcal{T}_t}
    \mathbf{H}^{(21)}_{\tau},
    \qquad
    \hat{\mathbf{h}}_t
    =
    \frac{\mathbf{h}_t}
         {\|\mathbf{h}_t\|_2},
\end{equation}

where $\mathcal{T}_t$ is the set of included token positions for turn $t$.

The normalized turn representation was cosine-projected onto each of the
23 denoised emotion directions:

\begin{equation}
    s_{t,e}
    =
    \hat{\mathbf{h}}_t
    \cdot
    \hat{\mathbf{v}}_e ,
    \qquad
    s_{t,e}\in[-1,1].
\end{equation}

Each turn therefore receives a 23-dimensional score vector

\begin{equation}
    \mathbf{s}_t
    =
    (s_{t,1},\ldots,s_{t,23})
    \in \mathbb{R}^{23}.
\end{equation}

Attorney and witness turns were encoded separately. The trajectory analyses
reported here use the witness-side behavioral trace.

\subsection{Temporal Normalization}
\label{app:trajectory-normalization}

Depositions vary substantially in length, preventing direct comparison of
their raw turn indices. We therefore normalized each deposition to relative
position within the examination.

For a deposition containing $T$ turns, the original turn positions were
mapped to the interval $[0,1]$ and linearly interpolated to $B=20$ equally
spaced temporal bins. For emotion $e$,

\begin{equation}
    \tilde{s}_{b,e}
    =
    \operatorname{interp}
    \left(
        \frac{b-1}{B-1};
        \frac{t-1}{T-1},
        s_{t,e}
    \right),
    \qquad
    b\in\{1,\ldots,B\}.
\end{equation}

Each deposition is therefore represented by an emotion-trajectory matrix

\begin{equation}
    \tilde{\mathbf{S}}
    \in
    \mathbb{R}^{20\times23}.
\end{equation}

For analyses requiring a single transcript-level representation, this matrix
was flattened to

\begin{equation}
    \mathbf{a}
    =
    \operatorname{vec}(\tilde{\mathbf{S}})
    \in
    \mathbb{R}^{460}.
\end{equation}

The use of 20 normalized bins trades temporal resolution for comparability
across examinations of substantially different lengths. The resulting arcs
therefore capture coarse longitudinal structure rather than turn-by-turn
correspondence.

\subsection{Mean Real--Synthetic Arc Similarity}
\label{app:mean-arc-similarity}

For each of the ten synthetic archetypes, we computed a mean synthetic
emotion arc by averaging the normalized trajectories across the ten
underlying witnesses. We likewise computed a mean real trajectory across
the real deposition corpus.

For each emotion and synthetic archetype, we calculated the Pearson
correlation between its 20-bin real and synthetic trajectories. The reported
overall correlation is the mean across the resulting emotion--archetype
comparisons:

\begin{equation}
    \bar{r}
    =
    \frac{1}{23\times10}
    \sum_{e=1}^{23}
    \sum_{a=1}^{10}
    r_{e,a}.
\end{equation}

The observed mean correlation was

\[
    \bar{r}=0.1477 \approx 0.148.
\]

The modest magnitude is important: this result indicates limited shared
temporal structure rather than close replication of real trajectories.

\subsection{Permutation Test}
\label{app:trajectory-permutation}

To determine whether the observed mean correlation could arise from similar
marginal emotion-score distributions without shared temporal ordering, we
constructed a permutation null distribution.

For each of 1,000 permutations, the 20 temporal bins of the synthetic
trajectories were randomly reordered while the real trajectories were left
unchanged. Correlations were then recomputed using the same procedure as for
the observed data. This preserves each synthetic trajectory's marginal score
distribution while destroying its temporal ordering.

The null distribution was centered near zero,

\[
    \mathrm{mean}(r_{\mathrm{null}})=0.0002,
\]

with a 95th percentile of

\[
    r_{\mathrm{null},.95}=0.0605.
\]

The observed value, $r=0.1477$, exceeded all 1,000 permuted correlations,
yielding an empirical $p<.001$.

\begin{figure}[t]
    \centering
    \includegraphics[width=0.65\textwidth]{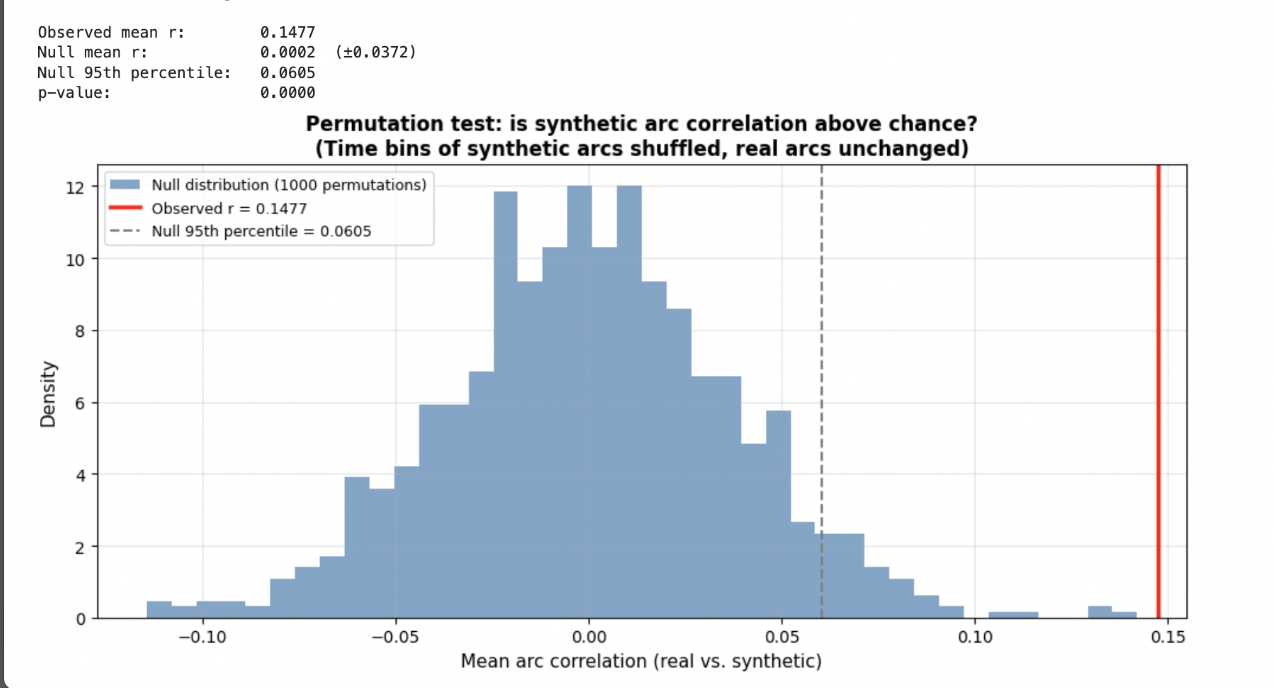}
    \caption{
    Null distribution for the trajectory-correlation analysis. Synthetic
    20-bin trajectories were temporally permuted 1,000 times while real
    trajectories were held fixed. The observed mean correlation
    ($r=0.1477$) exceeded all permuted values. The null mean was $0.0002$
    and its 95th percentile was $0.0605$.
    }
    \label{fig:trajectory-permutation}
\end{figure}

\subsection{Per-Witness Arc Similarity}
\label{app:per-witness-arcs}

The aggregate analysis does not indicate whether the same synthetic
archetype best approximates every real witness. We therefore additionally
performed a per-witness comparison.

For each of the ten witnesses, all available real deposition trajectories
for that witness were averaged to obtain

\[
    \tilde{\mathbf{S}}^{\mathrm{real}}_w
    \in
    \mathbb{R}^{20\times23}.
\]

This real arc was compared separately with each of the ten synthetic
archetype trajectories generated for the same underlying witness.

For emotion $e$, archetype $a$, and witness $w$, we computed

\begin{equation}
    r_{w,e,a}
    =
    \operatorname{PearsonR}
    \left(
        \tilde{s}^{\mathrm{real}}_{w,:,e},
        \tilde{s}^{\mathrm{syn}}_{w,:,e,a}
    \right).
\end{equation}

The best-fitting archetype for each witness was defined as the archetype
with the largest mean correlation across the 23 emotion dimensions:

\begin{equation}
    a_w^{*}
    =
    \arg\max_a
    \frac{1}{23}
    \sum_{e=1}^{23}
    r_{w,e,a}.
\end{equation}

For visualization, archetype columns are ordered by their mean correlation
and emotion rows by their variance across archetypes, placing the dimensions
that most strongly discriminate among simulated styles near the top.

For Jeffrey Kilper, the best-fitting synthetic trajectory was the
Cooperative archetype, with mean $r=.229$. As shown in the main text,
several rises and declines align qualitatively between the real and synthetic
trajectories, although the synthetic score range is substantially smaller.

The best-fitting archetypes for the ten witnesses were:

\begin{center}
\begin{tabular}{ll}
\toprule
\textbf{Witness} & \textbf{Best-fitting archetype} \\
\midrule
Jeffrey Kilper & cooperative \\
Catherine Jackson & inventive \\
Hugh O'Neill & inventive \\
Jane Williams & cooperative \\
John Adams & combative \\
Kirk Dumont & loquacious \\
Mark Pugh & nervous \\
Michael Wessler & dogmatic \\
Tiffany Kilper & combative \\
Todd Dean & dogmatic \\
\bottomrule
\end{tabular}
\end{center}

\subsection{Per-Witness Similarity Heatmaps}
\label{app:trajectory-heatmaps}

\begin{figure}[t]
\centering
\includegraphics[width=0.55\textwidth]{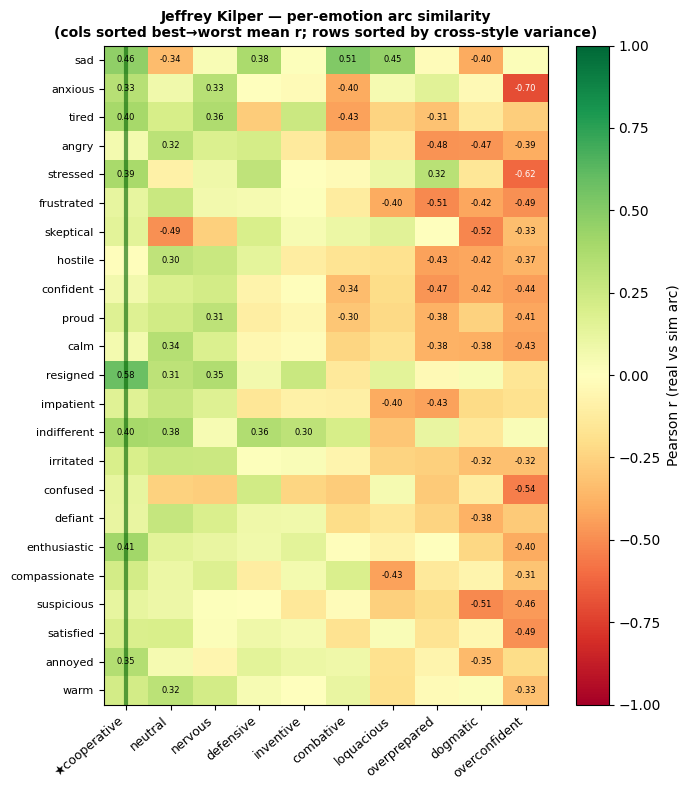}
\caption{Per-witness emotion-arc similarity matrix for Jeffrey Kilper.
The Cooperative archetype provides the best-fitting synthetic trajectory
($r=.229$).}
\label{fig:trajectory-heatmap-kilper}
\end{figure}

\begin{figure}[t]
\centering
\includegraphics[width=0.45\textwidth]{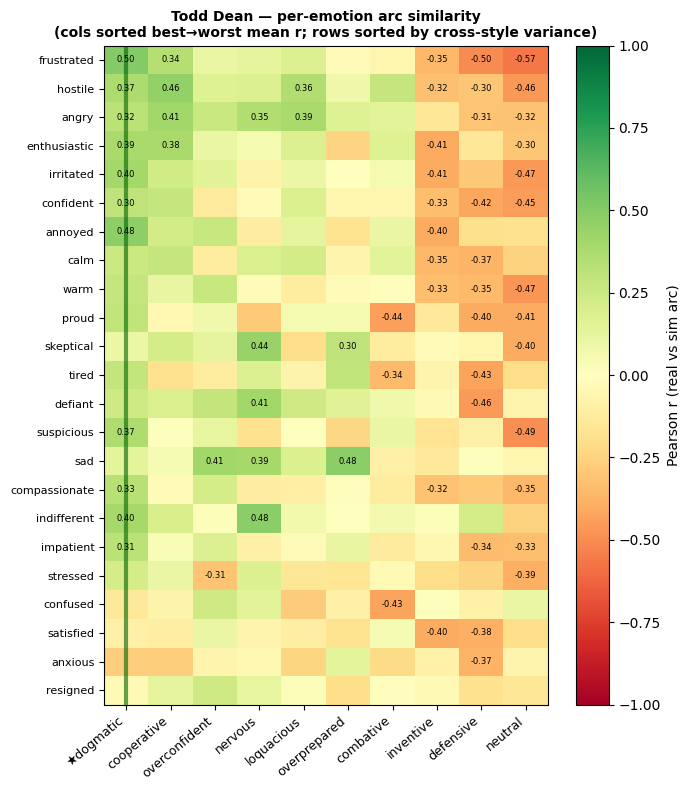}
\caption{Per-witness emotion-arc similarity matrix for Todd Dean;
best-fitting archetype: \textit{dogmatic}.}
\label{fig:trajectory-heatmaps-5}
\end{figure}

Figures~\ref{fig:trajectory-heatmap-kilper}--%
\ref{fig:trajectory-heatmaps-5} show the per-witness similarity matrices for
all ten witnesses. Each cell represents the Pearson correlation between a
real 20-bin emotion trajectory and the corresponding synthetic trajectory
for one archetype. Archetype columns are ordered by mean correlation across
the 23 emotion dimensions, and emotion rows are ordered by cross-archetype
variance.

\begin{figure}[t]
\centering
\begin{minipage}{0.45\textwidth}
    \centering
    \includegraphics[width=\textwidth]{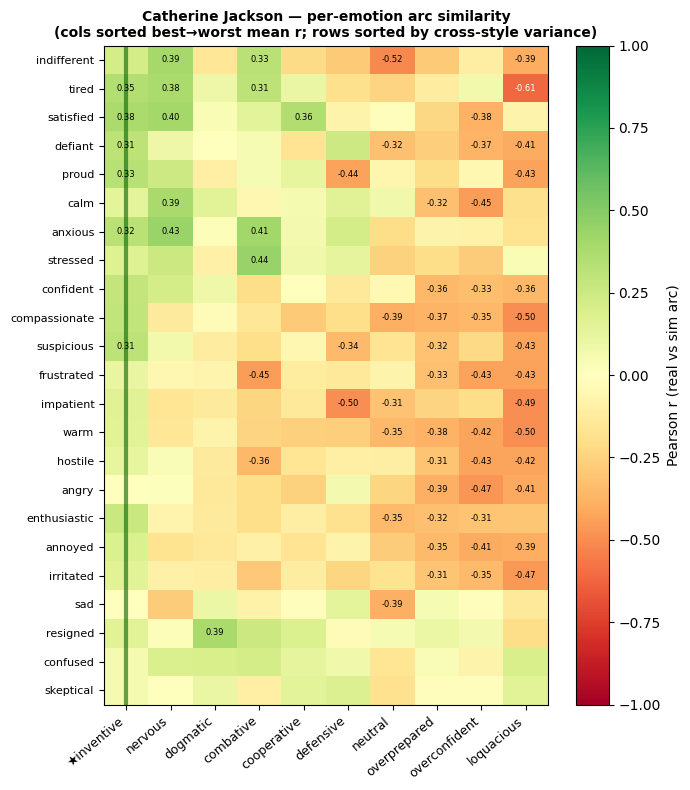}

    \small Catherine Jackson --- best fit: \textit{inventive}
\end{minipage}
\hfill
\begin{minipage}{0.45\textwidth}
    \centering
    \includegraphics[width=\textwidth]{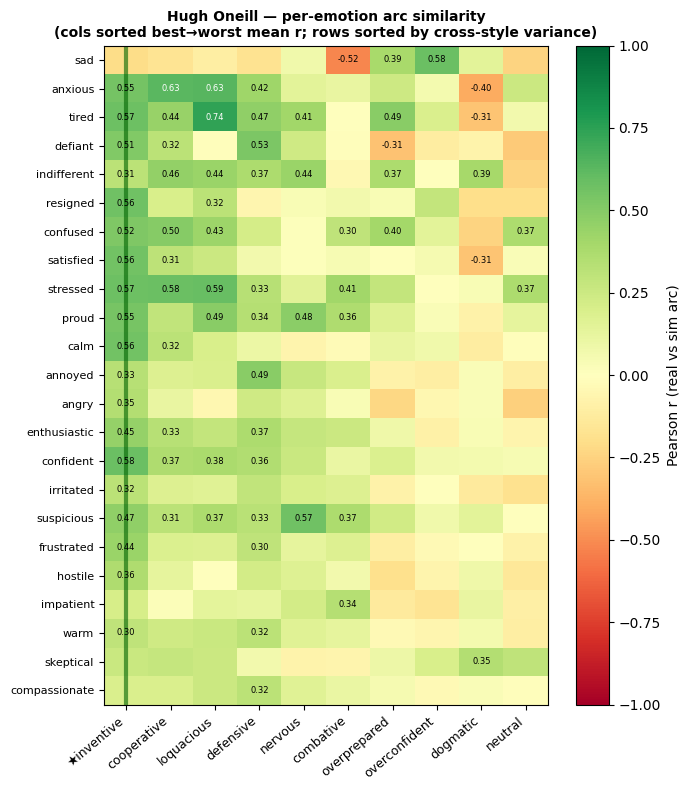}

    \small Hugh O'Neill --- best fit: \textit{inventive}
\end{minipage}
\caption{Per-witness emotion-arc similarity matrices.}
\label{fig:trajectory-heatmaps-1}
\end{figure}

\begin{figure}[t]
\centering
\begin{minipage}{0.45\textwidth}
    \centering
    \includegraphics[width=\textwidth]{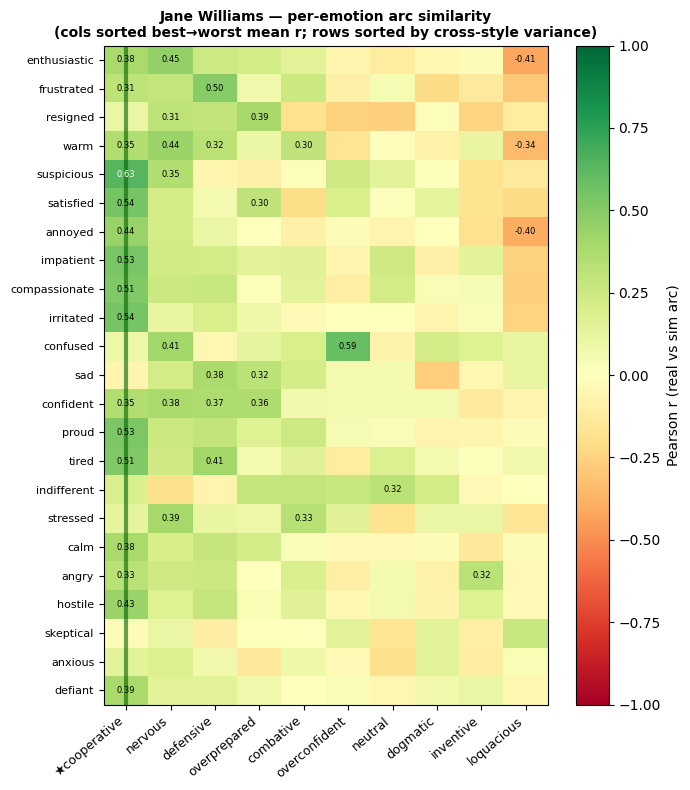}

    \small Jane Williams --- best fit: \textit{cooperative}
\end{minipage}
\hfill
\begin{minipage}{0.45\textwidth}
    \centering
    \includegraphics[width=\textwidth]{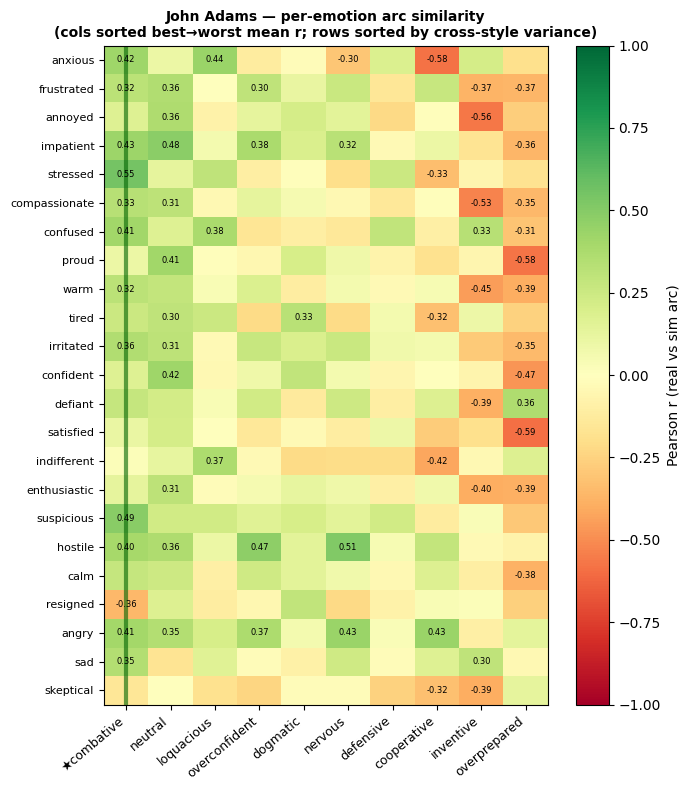}

    \small John Adams --- best fit: \textit{combative}
\end{minipage}
\caption{Per-witness emotion-arc similarity matrices.}
\label{fig:trajectory-heatmaps-2}
\end{figure}

\begin{figure}[t]
\centering
\begin{minipage}{0.45\textwidth}
    \centering
    \includegraphics[width=\textwidth]{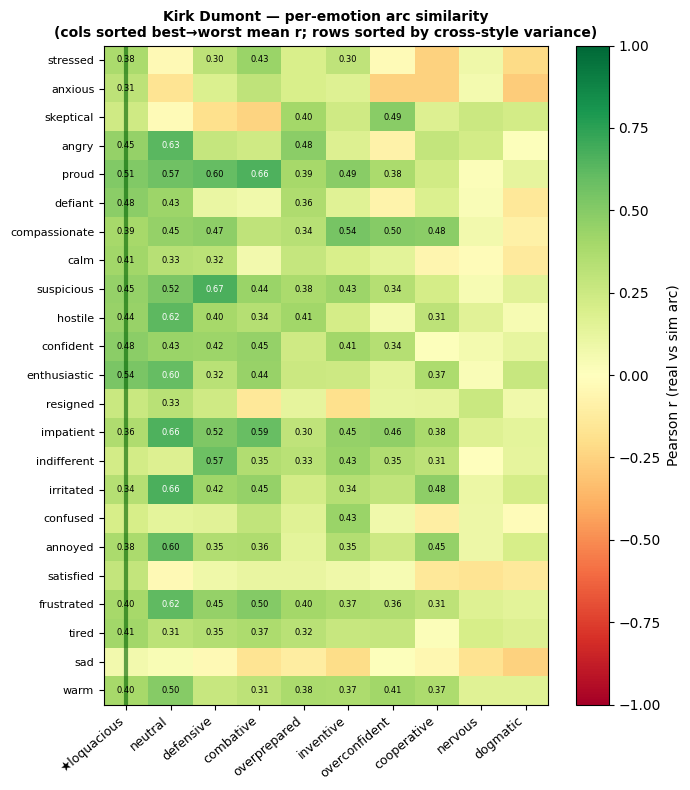}

    \small Kirk Dumont --- best fit: \textit{loquacious}
\end{minipage}
\hfill
\begin{minipage}{0.45\textwidth}
    \centering
    \includegraphics[width=\textwidth]{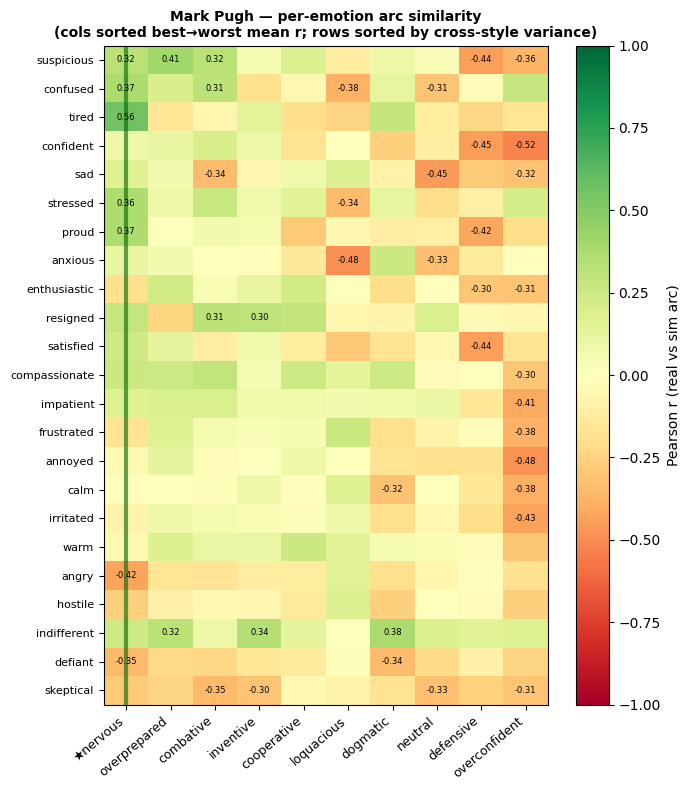}

    \small Mark Pugh --- best fit: \textit{nervous}
\end{minipage}
\caption{Per-witness emotion-arc similarity matrices.}
\label{fig:trajectory-heatmaps-3}
\end{figure}

\begin{figure}[t]
\centering
\begin{minipage}{0.45\textwidth}
    \centering
    \includegraphics[width=\textwidth]{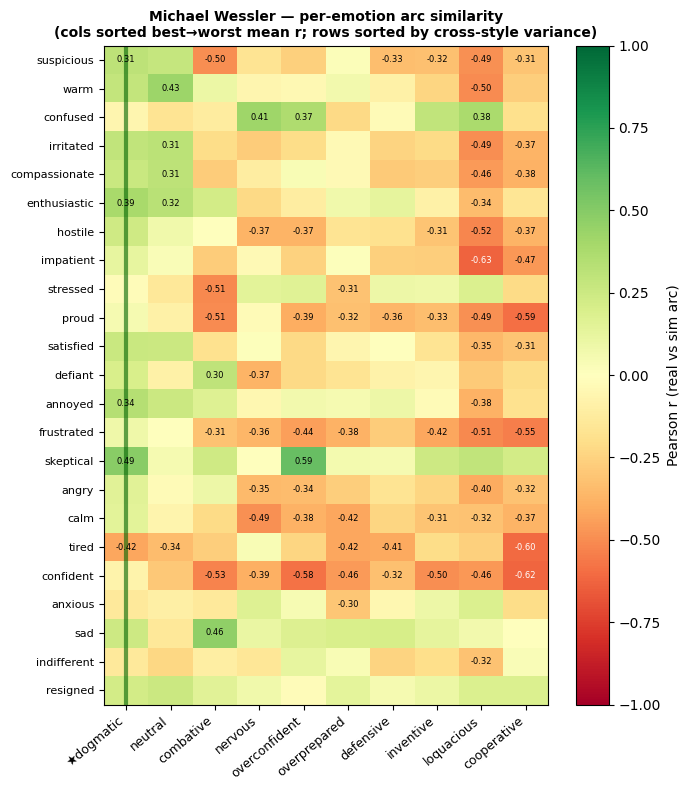}

    \small Michael Wessler --- best fit: \textit{dogmatic}
\end{minipage}
\hfill
\begin{minipage}{0.45\textwidth}
    \centering
    \includegraphics[width=\textwidth]{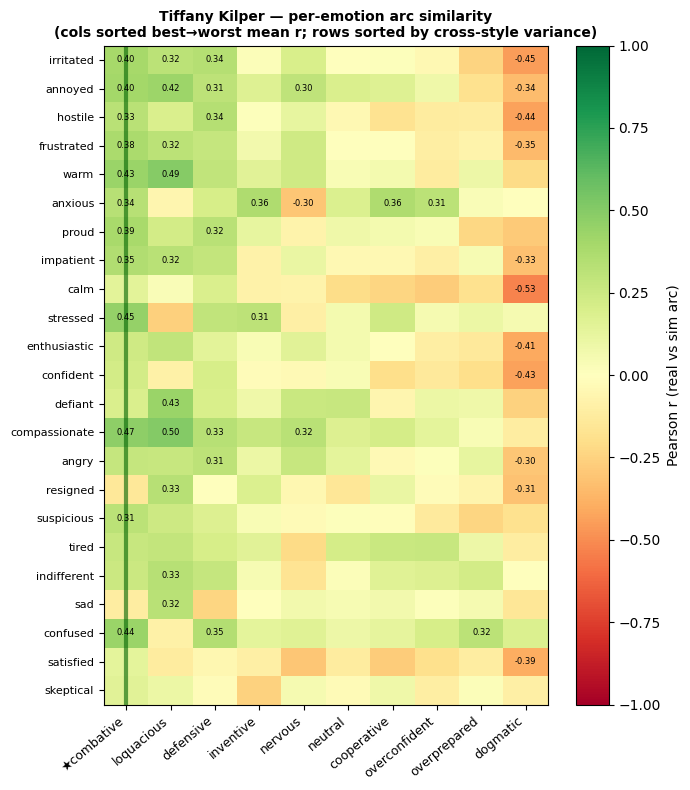}

    \small Tiffany Kilper --- best fit: \textit{combative}
\end{minipage}
\caption{Per-witness emotion-arc similarity matrices.}
\label{fig:trajectory-heatmaps-4}
\end{figure}

\subsection{Additional Per-Witness Trajectory Visualizations}
\label{app:trajectory-visualizations}

The main text shows the real and best-fitting synthetic trajectory for
Jeffrey Kilper. The following plots show the corresponding comparisons for
the remaining nine witnesses. These visualizations complement the correlation
statistics by illustrating both locally aligned temporal movements and the
generally smaller dynamic range of the synthetic trajectories.

\begin{figure}[t]
\centering
\begin{minipage}{0.48\textwidth}
\centering
\includegraphics[width=\linewidth, trim=0 0 2050 0, clip]{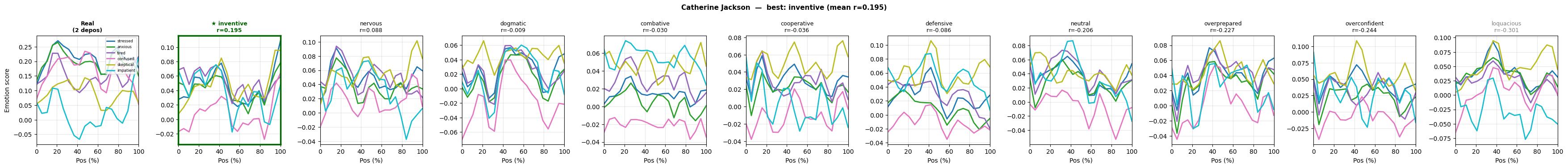}

\small Catherine Jackson --- best fit: \textit{inventive}
\end{minipage}
\hfill
\begin{minipage}{0.48\textwidth}
\centering
\includegraphics[width=\linewidth, trim=0 0 2060 0, clip]{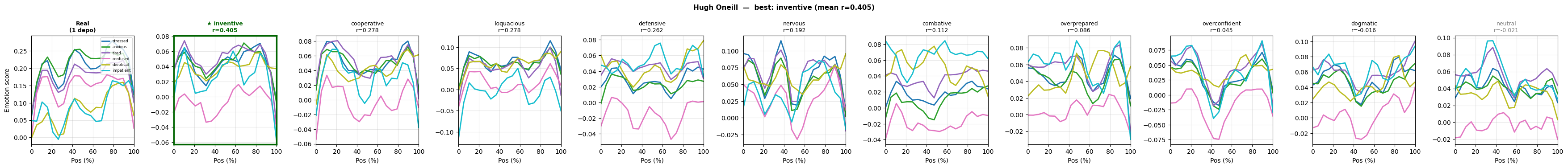}

\small Hugh O'Neill --- best fit: \textit{inventive}
\end{minipage}
\caption{Real and best-fitting synthetic emotion trajectories.}
\end{figure}

\begin{figure}[t]
\centering
\begin{minipage}{0.48\textwidth}
\centering
\includegraphics[width=\linewidth, trim=0 0 2050 0, clip]{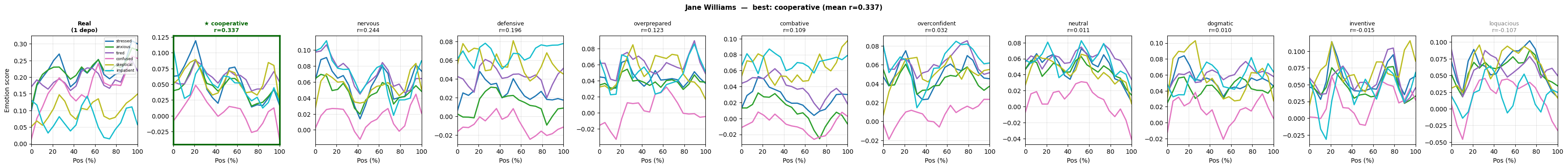}

\small Jane Williams --- best fit: \textit{cooperative}
\end{minipage}
\hfill
\begin{minipage}{0.48\textwidth}
\centering
\includegraphics[width=\linewidth, trim=0 0 2050 0, clip]{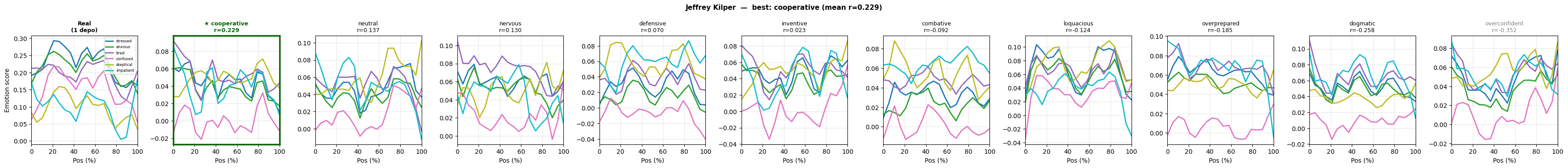}

\small John Adams --- best fit: \textit{combative}
\end{minipage}
\caption{Real and best-fitting synthetic emotion trajectories.}
\end{figure}

\begin{figure}[t]
\centering
\begin{minipage}{0.48\textwidth}
\centering
\includegraphics[width=\linewidth, trim=0 0 2060 0, clip]{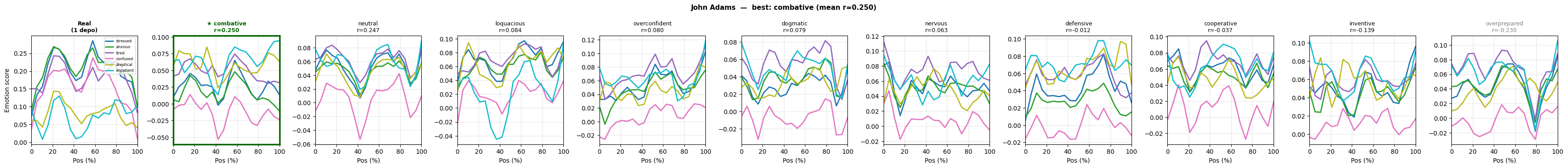}

\small Kirk Dumont --- best fit: \textit{loquacious}
\end{minipage}
\hfill
\begin{minipage}{0.48\textwidth}
\centering
\includegraphics[width=\linewidth, trim=0 0 2070 0, clip]{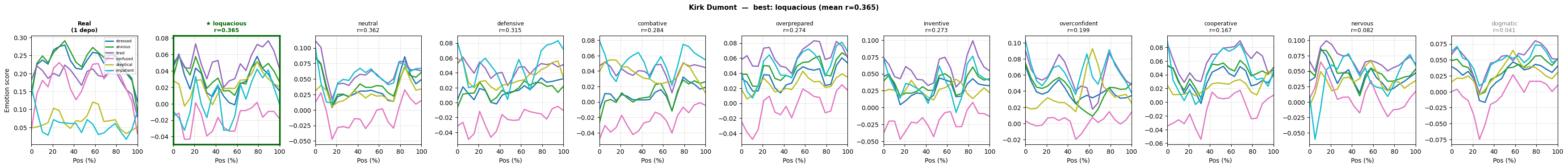}

\small Mark Pugh --- best fit: \textit{nervous}
\end{minipage}
\caption{Real and best-fitting synthetic emotion trajectories.}
\end{figure}

\begin{figure}[t]
\centering
\begin{minipage}{0.48\textwidth}
\centering
\includegraphics[width=\linewidth, trim=0 0 2060 0, clip]{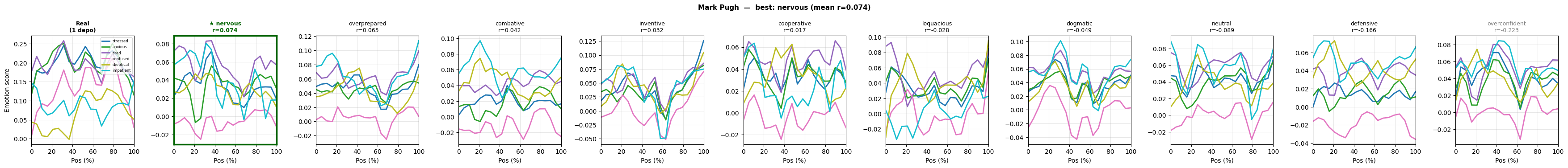}

\small Michael Wessler --- best fit: \textit{dogmatic}
\end{minipage}
\hfill
\begin{minipage}{0.48\textwidth}
\centering
\includegraphics[width=\linewidth, trim=0 0 2060 0, clip]{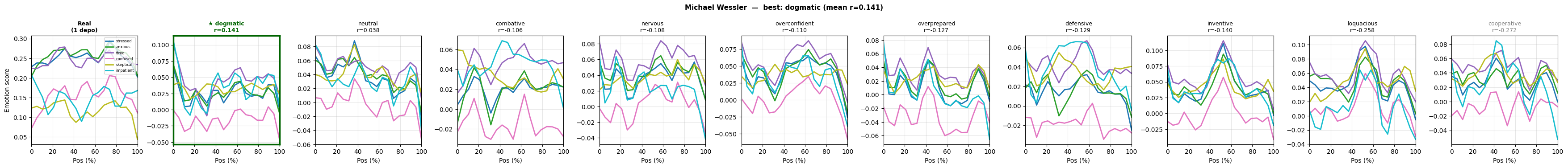}

\small Tiffany Kilper --- best fit: \textit{combative}
\end{minipage}
\caption{Real and best-fitting synthetic emotion trajectories.}
\end{figure}

\begin{figure}[t]
\centering
\includegraphics[
    width=0.48\textwidth,
    trim=0 0 2070 0,
    clip
]{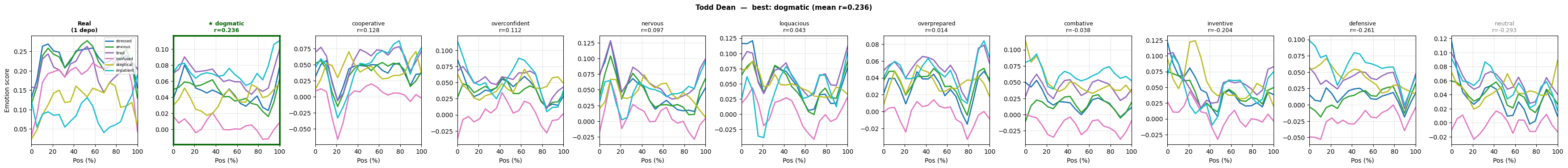}
\caption{Real and best-fitting synthetic emotion trajectory for Todd Dean
(\textit{dogmatic}).}
\end{figure}

\subsection{Magnitude Compression}
\label{app:trajectory-compression}

Although the temporal correlation analysis identifies above-chance shared
structure, the synthetic and real trajectories differ substantially in
magnitude. In the representative Jeffrey Kilper comparison, real emotion
scores span approximately $0.05$--$0.30$, whereas the synthetic scores span
approximately $0.00$--$0.08$.

The smaller synthetic range is also visible across the per-witness
trajectory plots. We therefore interpret the positive temporal correlation
as evidence of partial alignment in the \emph{direction and timing} of
behavioral change rather than matching emotional intensity. This compression
constitutes a systematic fidelity gap and provides a concrete target for
future simulator calibration.

\subsection{Real--Synthetic PCA}
\label{app:trajectory-pca}

To test whether real and synthetic transcripts occupy similar regions of the
trajectory representation space, each normalized
$20\times23$ trajectory was flattened to a 460-dimensional vector.
Features were standardized before applying principal component analysis.

The first principal component explained 29.9\% of total variance. As shown
in Figure~\ref{fig:trajectory-pca} in the main text, synthetic transcripts
clustered relatively tightly in one region of the embedding, whereas real
transcripts were substantially more dispersed and largely separated along
PC1.

This separation provides an important counterpoint to the positive
correlation result. The two analyses measure different properties:
correlation tests whether trajectories share temporal rises and declines,
whereas PCA captures broader multivariate differences in magnitude,
dispersion, and covariance structure. Thus, above-chance temporal alignment
does not imply distributional equivalence between real and synthetic
depositions.

\subsection{Story-Generation Topics and Prompt}
\label{app:emotion-story-prompt}

For completeness, the 25 topics used to construct the emotion directions
are listed below. Using multiple unrelated topics reduces the likelihood
that an emotion direction primarily captures one recurring semantic domain.

\begin{multicols}{2}
\begin{enumerate}
    \item A person discovers their old love letters have been turned into a stage play.
    \item A coworker quietly takes credit for a group project during a company meeting.
    \item A parent finds out their child has been skipping therapy sessions.
    \item Someone learns their identical twin has been impersonating them online.
    \item A person receives a voicemail meant for someone who died.
    \item A gardener finds a neighbor has been harvesting their vegetable garden.
    \item A person discovers their estranged sibling lives two streets away.
    \item Someone finds out their closest friend testified against them without telling them.
    \item A person's dog recognizes a stranger in a way that suggests a prior connection.
    \item A teenager finds their parent's old arrest record.
    \item A person receives a corrected version of their own memory from a therapist's notes.
    \item Someone learns their wedding venue has been double-booked.
    \item A person discovers their child has been secretly supporting a distant relative.
    \item An employee finds out their resignation letter was never submitted by their manager.
    \item A person learns the eulogy they wrote for a funeral was significantly rewritten.
    \item Someone finds a photograph of themselves at a place they have no memory of visiting.
    \item A person's handwritten recipe is printed on a mass-produced product without credit.
    \item A coworker confesses they have been covering for a mutual colleague's absences.
    \item Someone discovers their landlord has been entering the apartment unannounced.
    \item A parent learns their adult child has quietly paid off a family debt.
    \item A person realizes their therapist and their boss know each other socially.
    \item Someone finds their name listed as a dedication in a stranger's published memoir.
    \item A person learns their childhood nickname became a running joke among relatives.
    \item Two old friends realize they were in the same hospital on the same night years ago.
    \item A person discovers their long-term pen pal has been using a fictitious name.
\end{enumerate}
\end{multicols}

The story-generation prompt was:

\begin{quote}
\ttfamily\small

Write \{n\_stories\} different stories based on the following premise.

\medskip
Topic: \{topic\}

\medskip
The story should follow a character who is feeling \{emotion\}.

\medskip
Format the stories like so:

\medskip
[story 1]

[story 2]

[story 3]

\medskip
etc.

\medskip
The paragraphs should each be a fresh start, with no continuity. Try to
make them diverse and not use the same turns of phrase. Across the different
stories, use a mix of third-person narration and first-person narration.

\medskip
IMPORTANT: You must NEVER use the word `\{emotion\}' or any direct synonyms
of it in the stories. Instead, convey the emotion ONLY through:

\begin{itemize}
    \item The character's actions and behaviors
    \item Physical sensations and body language
    \item Dialogue and tone of voice
    \item Thoughts and internal reactions
    \item Situational context and environmental descriptions
\end{itemize}

The emotion should be clearly conveyed to the reader through these indirect
means, but never explicitly named.
\end{quote}

\subsection{Interpretive Scope and Limitations}
\label{app:trajectory-limitations}

The emotion-vector analysis is exploratory and should be interpreted as a
proxy for longitudinal affective and interpersonal structure rather than as
a validated measure of witness emotion. In particular, the emotion
directions are derived from language-model representations of generated
narratives, not from ground-truth psychological labels on deposition
testimony.

The extraction layer is also inherited from prior methodology rather than
optimized specifically for this deposition corpus. Layer 21 was selected as
a mid-to-late residual-stream representation, but the optimal layer may vary
across model architectures and domains.

Temporal normalization introduces a further limitation. Interpolating each
deposition to 20 bins makes examinations of different lengths directly
comparable, but necessarily removes fine-grained timing information and
treats equivalent relative positions as comparable even when the underlying
questioning structure differs.

Finally, the real corpus contains only 55 transcripts from ten witnesses in
a single multidistrict litigation. The trajectory results therefore provide
a proof of concept for longitudinal behavioral evaluation rather than a
general estimate of real--synthetic behavioral fidelity across legal
domains.
\end{document}